\documentclass[journal,twoside]{IEEEtran}
\usepackage{amssymb}
\usepackage{amsthm}
\usepackage{multirow}
\usepackage{inputenc}
\usepackage{enumerate} 
\usepackage{textcomp}
\usepackage{listings}
\usepackage{array}
\usepackage[switch,mathlines]{lineno}
\usepackage{soul}
\usepackage{xfrac}
\usepackage{graphicx}
\usepackage{cite}
\usepackage[cmex10]{amsmath}
\usepackage{algorithm}
\usepackage{xcolor}
\usepackage{colortbl}

\usepackage{cancel}

\usepackage{algorithm,algpseudocode}
\algnewcommand{\Inputs}[1]{%
  \State \textbf{Inputs:}
  \Statex \hspace*{\algorithmicindent}\parbox[t]{.8\linewidth}{\raggedright #1}
}
\algnewcommand{\Initialize}[1]{%
  \State \textbf{Initialize:}
  \Statex \hspace*{\algorithmicindent}\parbox[t]{.8\linewidth}{\raggedright #1}
}

\makeatletter
\def\footnoterule{\kern-3\p@
  \hrule \@width 3.3in \kern 2.6\p@} 
\makeatother

\usepackage{algpseudocode}

\usepackage{stackengine}

\usepackage{algpseudocode}
\usepackage{CJK}
\usepackage[cmex10]{amsmath}
\usepackage{bm}
\usepackage{color}
\usepackage{amsmath,amsthm,amssymb,amsfonts}
\usepackage{flushend}
\usepackage{float}

\newtheorem{remark}{Remark}

\usepackage{arydshln} 
\usepackage{hyperref}
\hypersetup{
     colorlinks   = true,
     linkcolor    = blue,
     citecolor    = red,
     urlcolor     = blue
}
\makeatletter
\newcommand*{\transpose}{%
  {\mathpalette\@transpose{}}%
}
\newcommand*{\@transpose}[2]{%
  \raisebox{\depth}{$\m@th#1\intercal$}%
}
\makeatother

\usepackage{soul}
\usepackage{tikz}
\usepackage{booktabs}
\usepackage{tabularx}

\usepackage[caption=false,font=footnotesize]{subfig}
\ifCLASSINFOpdf
\else
\fi

\begin{document}
\renewcommand{\ttdefault}{cmtt}
\bstctlcite{IEEEexample:BSTcontrol}

\title{An Adaptive-Importance-Sampling-Enhanced Bayesian Approach for Topology Estimation in an Unbalanced Power Distribution System}



\author{
{Yijun~Xu,~\IEEEmembership{Senior Member}, 
Jaber Valinejad,~\IEEEmembership{Student Member},
Mert~Korkali,~\IEEEmembership{Senior Member},
Lamine~Mili,~\IEEEmembership{Life Fellow},
Yajun Wang,~\IEEEmembership{Senior Member},
Xiao~Chen,}
Zongsheng Zheng,~\IEEEmembership{Member} 
       \thanks{Y. Xu, J. Valinejad and L. Mili  are with the Bradley Department of Electrical and Computer Engineering, Virginia Tech, Northern Virginia Center, Falls Church, VA 22043 USA (e-mail:\{yijunxu,jabervalinejad, lmili\}@vt.edu).}
\thanks{M.~Korkali is with the Computational Engineering Division, Lawrence Livermore National Laboratory, Livermore, CA 94550 USA (e-mail: korkali1@llnl.gov).}
\thanks{Y.~Wang is with the Distribution Grid Solutions, Dominion Energy Virginia, Richmond, VA 23219 USA (e-mail: yajun.wang@dominionenergy.com).}
\thanks{X.~Chen is with the Center for Applied Scientific Computing, Lawrence Livermore National Laboratory, Livermore, CA 94550 USA (e-mail: chen73@llnl.gov).}
 	\thanks{Z.~Zheng is with the College of Electrical Engineering, Sichuan University, Chengdu, 610065 China (e-mail: zongsheng56@126.com).}
\thanks{This work was supported, in part, by the U.S. National Science Foundation under  Grant 1917308, by the Scientific Research Startup Fund for Introducing Talents of Sichuan University under Grant 1082204112576, and by the United States Department of Energy Office of Electricity Advanced Grid Modeling (AGM) Program, and performed under the auspices of the U.S. Department of Energy by Lawrence Livermore National Laboratory under Contract DE-AC52-07NA27344. 
Document released as LLNL-JRNL-820867.
}
}

\markboth{IEEE Transactions on Power Systems}%
{Xu \MakeLowercase{\textit{\textit{et al.}}}: An Adaptive-Importance-Sampling-Enhanced Bayesian Approach for Topology Estimation in a Distribution System}
\maketitle
\begin{abstract}
The reliable operation of a power distribution system relies on a good prior knowledge of its topology and its system state. Although crucial, due to the lack of direct monitoring devices on the switch statuses, the topology information is often unavailable or outdated for the distribution system operators for real-time applications. Apart from the limited observability of the power distribution system, other challenges are the nonlinearity of the model, the complicated, unbalanced structure of the distribution system, and the scale of the system. To overcome the above challenges, this paper proposes a Bayesian-inference framework that allows us to simultaneously estimate the topology and the state of a three-phase, unbalanced power distribution system. Specifically, by using the very limited number of measurements available that are associated with the forecast load data,  we  efficiently recover the full Bayesian posterior distributions of the system topology under both normal and outage operation conditions.  This is performed through an adaptive importance sampling procedure that greatly alleviates the computational burden of the traditional Monte-Carlo (MC)-sampling-based approach while maintaining a good estimation accuracy.  The simulations conducted on the IEEE 123-bus test system and an unbalanced  1282-bus system reveal the excellent performances of the proposed method. 
\end{abstract}


\IEEEpeerreviewmaketitle

\section{Introduction}
\IEEEPARstart{A}{n accurate} and efficient estimation of the topology in power distribution systems is becoming an important and timely research subject. On one hand, it serves as a prerequisite for the reliable and efficient operations and plannings of modern power distribution systems where the deployment of the renewables and other distributed energy resources are increasing rapidly. On the other hand, it serves as a fundamental tool for a  fast restoration of the power distribution system after an unexpected disruptive event. However, the network topology is typically unavailable or outdated due to the limited information of the network switch statuses and their insufficient visual verification by crew members. Furthermore, the access to switch statuses of the underground cables in urban areas by crew members can be costly, time-consuming, and labor-intensive, which makes it impossible to rely on for online applications. 

Although the topology estimation problems have been studied extensively in the past decades for power systems, mature techniques are, in general, more focused on transmission systems. Some examples include a robust Huber estimator proposed by Mili \emph{et al.} \cite{mili1999robust} and a Bayesian-based hypothesis testing advocated by Louren\c co \emph{et al.} \cite{lourenco2004bayesian} to identify the topology errors; a traveling-wave-based technique initiated by Korkali and Abur \cite{korkali2011traveling,korkali2013robust} to locate the source of topology change caused by disturbances; a mixed-integer quadratic programming considered by Caro \emph{et al.} \cite{caro2009breaker} to estimate the switch status.  As for the simultaneous detection of multiple outages, an offline-trained model is suggested by Zhao \emph{et al.}  \cite{zhao2019learning}
{Besides, 
an interesting state estimation procedure without topology processor was advocated by Donmez and Abur \cite{donmez2021state} and a neural-network-based approach was  introduced by Krstulovic \emph{et al.} \cite{krstulovic2013towards}}, to cite a few. Yet, considering that the distribution system is, typically, radially operated in an unbalanced three-phase structure with low observability, these techniques cannot be naturally extended to the distribution systems  \cite{gandluru2019joint}. 

To overcome the above difficulties, more and more researches are conducted in the power distribution system topology estimation today. Apart from the literature focusing solely on the general grid structure learning~\cite{deka2019topology,deka2017structure}, or the switch statuses for the reconfiguration tracking~\cite{cavraro2019real}, some research studies also explore the joint estimation considering the topology uncertainty as follows. More specifically, Deka~\emph{et al.} \cite{deka2020joint} propose to utilize a spanning-tree-based graphical model to jointly estimate the topology and the power injections. Similarly, the topology and the line parameter joint estimation are explored by Park~\emph{et al.}  \cite{park2020learning} using graph theory, and by Yu \emph{et al.}  \cite{yu2017patopa,yu2018patopaem} using a data-driven approach, etc. Besides, a topology and outage joint estimation scheme is recently addressed by Gandluru \emph{et al.} \cite{gandluru2019joint}. Typically, the abovementioned distribution system topology estimation related works have the following concerns.  First, the distribution system model is complicated due to its nonlinear model and unbalanced three-phase structure. To alleviate these difficulties, some researches adopt a DC model \cite{kekatos2015online,zhao2019learning,sevlian2015distribution} or a linearized model \cite{deka2017structure,gandluru2019joint,cavraro2019real,baldwin1993power}.  Some works simplify the three-phase structure by ignoring the mutual coupling between phases with a single-phase distribution system model~\cite{deka2017structure,yu2017patopa,yu2018patopaem,cavraro2019real,bhela2017enhancing}. Although simple and straightforward, all these simplifications inevitably sacrifice model accuracy.  Second, the observability in the distribution system is typically insufficient since the utility operators monitor distribution grid with meters only at a few buses\cite{bhela2017enhancing}. Although the deployment of  measurement devices (e.g., phasor measurement units (PMUs)) is growing, it might still be bold to assume that all the buses are directly measured \cite{ardakanian2019identification}. Subsequently, optimal sensor placement is extensively studied in \cite{cavraro2019real,sevlian2017outage,baldwin1993power,gou2008generalized, xygkis2016fisher,liu2014optimal,liu2012trade}. Alternatively, different measurement devices are also advocated in the literature, such as smart meters \cite{bhela2017enhancing}, probing technique \cite{bhela2019smart}, ping measurement \cite{gandluru2019joint}, or even pseudomeasurement from forecast and historical data \cite{gandluru2019joint,liao2016urban}. Besides, using some publicly available market data (e.g., online energy prices) to enable topology tracking is innovatively proposed by Kekatos \emph{et al.} in \cite{kekatos2015online}. Third, the distribution system is, in general, radial. Even though exceptions exist for some mesh and loop structure in the urban area \cite{
liao2016urban}, topology changes in the distribution system are still likely to occur, most likely to induce outages. Therefore, it comes as no surprise that the recent topology estimation work is simultaneously conducted with outage estimation \cite{gandluru2019joint}. Lastly, the size of the distribution system remains a bottleneck for most of the existing methods, which are tested on small-size systems (e.g., the IEEE 13-bus and 33-bus systems \cite{ardakanian2019identification,deka2017structure,deka2019topology,singh2010recursive}), and the demonstration of methods on a utility large-scale system has not yet been fully explored \cite{gandluru2019joint}.

Facing these challenges, this paper proposes a new adaptive importance sampling (AIS) scheme under the Bayesian inference framework to simultaneously estimate the topology, the outages, and the power injections of a distribution system, while considering the latter as byproducts.
{Unlike the Bayesian topology inference proposed in~\cite{singh2010recursive}, our AIS-based Bayesian inference is \emph{derivative-free} and, therefore, can easily be extended to more complex, three-phase, unbalanced, larger-scale distribution systems. 
}

The contributions of this paper are as follows:
\begin{itemize} 
\item A formulation of a Bayesian-inference framework that enables a general operational topology, outage, and states joint estimation is provided. This framework has no limitation on the type of the model, which therefore, makes it applicable to a realistic nonlinear distribution system model with a three-phase unbalanced structure.
\item This Bayesian framework is further merged into a two-stage estimation procedure that enables us to not only use limited measurement (i.e., a meter in the primary feeder combined with meters only in small portion of the user-end and forecast data), but also theoretically eliminate the estimation bias caused by the incorrect pseudomeasurement, i.e., the forecast data, in the outage area without using any ping measurement for connectivity identification \cite{gandluru2019joint}.

\item  To avoid an exhaustive search of all possible topologies, which might be impractical for an online application to a large-scale system\cite{zamani2015topology,sevlian2015distribution,yu2017patopa}, we propose to merge an AIS scheme into the Bayesian-inference framework \cite{bugallo2017adaptive}, for the first time, in the distribution system topology estimation procedure to achieve a faster convergence with the adaptively fine-tuned parameter space. The weights of the AIS further facilitate the recovery of the Bayesian posteriors that quantify the confidence of the estimation. {It not only overcomes the drawbacks of the exhaustive search algorithm, but also outperforms the traditional importance sampling algorithm in terms of computing time and performance, and, therefore, can serve as a cost-effective tool for online applications.}
\end{itemize}

The performance of our proposed method has been analyzed through simulations that are carried out on an IEEE test feeder and a real utility-scale system. These simulations reveal the excellent performance of the proposed method from the standpoint of simulation accuracy and computing efficiency. {We also demonstrate that the proposed method has a quite stable performance for radial and loop-structured networks and for a wide range of R/X ratio, from moderate to very large values, which induce a strong nonlinearity of the model.}

This paper is organized as follows: in Section II, the problem formulation is presented. In Section III, the background on importance sampling and adaptive importance sampling are introduced. Section IV presents the proposed method. Case studies are presented in Section V, followed by the conclusions and future work in Section VI.

\vspace{-0.2cm}
\section{Problem Formulation}
In this section, we will first briefly introduce the three-phase power distribution system model. Then, we will also formulate it into the Bayesian inference framework.
\vspace{-0.2cm}
\subsection{Model Description}
\subsubsection{Basics of a Three-phase Distribution System}Following the notations in\cite{ardakanian2019identification}, let us use a graph $\mathcal{G}=(\mathcal{N},\mathcal{E})$ to represent a multi-phase power distribution system model. Here, the nodes, $\mathcal{N}=\{1,2,\dots,N\}$, corresponding to the $N$ buses and the edges, $\mathcal{E}\subseteq \mathcal{N}\times\mathcal{N}$, represents the set of the distribution lines. Each line connects ordered pair of Buses $(m, n)$ between Buses $m$ and $n$. To extend the notations into the distribution system, let $\mathcal{P}_{n}=\{a_{n}, b_{n}, c_{n}\}$ denote the three phases of the system at Bus $n$ and let $\mathcal{P}_{m, n}=\{a_{m, n}, b_{m, n}, c_{m, n}\}$ denote the phases of Line $(m, n)$.  Accordingly, we obtain the three-phase voltage with respect to ground at Bus $n$ as $V_{n}=\{{V_{n}}^{\phi}\}_{{\phi}\in\{a_{n}, b_{n}, c_{n}\}}$, and its injected currents, $I_{n}$, as $I_{n}=\{{I_{n}}^{\phi}\}_{{\phi}\in\{a_{n}, b_{n}, c_{n}\}}$, respectively. The current for Line $(m, n)$ is denoted as $I_{m, n}=\{{I_{m, n}}^{\phi}\}_{{\phi}\in\{a_{m, n}, b_{m, n}, c_{m, n}\}}$. 

Further, by denoting the phase-impedance and shunt-admittance matrices of the $\pi$-equivalent model, $(m,n)$, as $Z_{m, n}\in\mathbb{C}^{|\mathcal{P}_{m, n}|\times|\mathcal{P}_{m, n}|} $ and 
$Y_{m, n}\in\mathbb{C}^{|\mathcal{P}_{m, n}|\times|\mathcal{P}_{m, n}|}$,
where $\mathbb{C}$ represents the set of the complex matrices, and by considering the admittance matrices of all the other components (e.g., transformers), we obtain the assembled admittance matrix for the distribution system as $Y_\text{bus}$. Subsequently, we obtain
\begin{equation} \underbrace{{\left[\begin{array}{c}I_1 \\ I_2 \\ \vdots \\ I_N  \end{array}\right]}}_{I_{\text{bus}}}= \underbrace{{\left[\begin{array}{cccc}Y_{11} & Y_{12} & \ldots & Y_{1N} \\ Y_{21} & Y_{22} & \ldots & Y_{2N} \\ \vdots & \vdots & \ddots & \vdots \\ Y_{N1} & Y_{N2} & \ldots & Y_{NN} \end{array}\right]}}_{Y_{\text{bus}}} \underbrace{{\left[\begin{array}{c}V_1 \\ V_2 \\ \vdots \\ V_N  \end{array}\right]}}_{V_{\text{bus}}}.
\end{equation}

Till now, we have presented the basic model of a three-phase distribution system.  Within this framework, other system variables, such as the per-phase net  active  and  reactive  power  injection at Bus $n$, denoted by $P_{n}=\{{P_{n}}^{\phi}\}_{{\phi}\in\mathcal{P}_{n}}$ and $Q_{n}=\{{Q_{n}}^{\phi}\}_{{\phi}\in\mathcal{P}_{n}}$ respectively, and  the per-phase active  and  reactive  power  flow in the line connecting Buses $m$ and $n$, denoted by $P_{m, n}=\{{P_{m, n}}^{\phi}\}_{{\phi}\in\mathcal{P}_{m, n}}$ and $Q_{m, n}=\{{Q_{m, n}}^{\phi}\}_{{\phi}\in\mathcal{P}_{m, n}}$ respectively, can be calculated. The vectors that collect $P_{n}$ and $Q_{n}$ at a bus are defined as $\bm{P}_{b}$ and $\bm{Q}_{b}$, whereas the vectors that collect $P_{m, n}$ and $Q_{m, n}$ in a line are denoted as $\bm{P}_{l}$ and $\bm{Q}_{l}$, accordingly. 
\subsubsection{Switch Status in the Model}
Now, let us consider the topology uncertainty brought by the switch statuses. Following the existing literature \cite{cavraro2019real,gandluru2019joint,zhao2019learning}, let us introduce the binary variables capturing the status of the switch between Buses $m$ and $n$ as $s_{m, n}$. Here, let us define that $s_{m, n}=1$ if Line $(m,n)$ is connected and $s_{m, n}=0$ for a disconnected one. For a power distribution system with $N_s$ switches, we define the set of the switches as $\mathcal{S}$, and a vector $\bm{B}$ that collects the binary variables, $B_{{s}_{m, n}}$, to capture the status of the switch ${s}_{m, n}$.
Now, given switch statuses in a distribution system, we can recover its topology as reflected in the admittance matrix $Y_\text{bus}$ that can
be determined accordingly with the physical parameters of
the system. Till now, we have completed the presentation of the distribution system model considering its topology uncertainty. 

\begin{remark}
 {It is worth pointing out that the $Y_\text{bus}$-matrix elements are determined by the line parameters, the tap ratios of transformers or regulators, the capacitor banks as well as the switch statuses, among others. In this paper, we are interested in solving the topology uncertainty problem raised by the switch statuses as done in~\cite{zhao2019learning,gandluru2019joint}, and \cite{cavraro2019real}. The detailed parameter estimation problems for other model components are beyond the scope of this paper. In practice, although some parameters of the power distribution system model (e.g., the line parameters) may be unknown or not be precisely known, utilities can conduct parameter estimation as a prior stage of topology and state estimation ~\cite{gandluru2019joint}, which allows the latter tasks to be executed using a reasonable model. For example, Yu $et$ $al.$ in \cite{yu2017patopa} and \cite{yu2018patopaem} propose such an effective parameter-topology-parameter  joint estimation scheme.  In the same vein, parameter estimation of other model components (e.g., transformer tap ratios and line phases) can be performed  independently \cite{peppanen2016distribution, pires2013constrained, short2012advanced}.
 }
\end{remark}

\vspace{-0.5cm}
\subsection{Bayesian Inference}
Let us briefly introduce the Bayesian inference that has been widely applied in inverse problems in many industrial applications~\cite{kaipio2006statistical,tarantola2005inverse,xu2019bayesian,xu2019adaptive,petra2016bayesian}. 
First, following the notions in 
\cite{kaipio2006statistical}, let us express the forward model used in Bayesian inference as
\begin{equation}
    \bm{y}=\bm{f}(\bm{x})+\bm{e},
\end{equation}
where $\bm{y}\in \mathbb{R}^{D}$ contains the observations of dimension $D$; $\bm{x}\in \mathbb{R}^{N_{x}}$ is expressed as a random vector that contains the parameters to be estimated; $N_{x}$ is the number of parameters to be estimated, which depends on the specific application; $\bm{f(\cdot)}$ is the vector-valued forward function that includes the abovementioned distribution system model, which maps the model parameter vector $\bm{x}$ to the observation vector $\bm{y}$; $\bm{e}\in \mathbb{R}^{D}$ stands for the measurement-error vector whose components are assumed to be mutually independent random variables with the joint probability density functions (pdfs), $\pi_{\bm{e}}$, defined as
\begin{equation}
\label{memodel}
 \pi_{\bm{e}}=\prod_{i=1}^{D}\pi_{e_{i}}(e_{i}).   
\end{equation}
In the Bayesian inference, each parameter $x_{i}$ is also viewed as a random variable with a given prior probability distribution, whose pdf is denoted by $\pi_{i}(x_{i})$. Note that here $\bm{e}$ and $\bm{x}$ are assumed to be  mutually independent. 
The corresponding joint prior density function for a vector $\bm{x}$ is given by
$
    \label{priorm}
    \pi_{\text{prior}}(\bm{x})=\prod_{i=1}^{N_{x}}{\pi_{i}(x_{i})}.
$
Given the observation vector $\bm{y}$, the posterior pdf, $\pi_{\text{post}}(\bm{x}|\bm{y})$, for the parameter vector $\bm{x}$ is derived as 
\begin{equation}
\label{eq:post}
    \pi_{\text{post}}(\bm{x}|\bm{y})\propto{\pi_{\text{like}}(\bm{y|x}){\pi_{\text{prior}}(\bm{x})}},
\end{equation}
where $\pi_{\text{like}}(\bm{y|x})$ denotes the likelihood function expressed as
$
\label{likelihood}
  \pi_{\text{like}}(\bm{y|x})=\pi_{\bm{e}}(\bm{y}-\bm{f}(\bm{x})).
$
Apart from the full Bayesian posterior distribution, $\pi_{\text{post}}(\bm{x}|\bm{y})$, that allows us to quantify of the uncertainty of the unknown parameters, we utilize a vector of the deterministic value using a \emph{maximum-a-posteriori} (MAP) estimator defined as
\begin{equation}
\label{MAP}
\begin{aligned}
& \bm{\hat{x}}_{\text{MAP}}=\underset{\bm{x}}{\text{arg min}}
&  \{-\pi_{\text{post}}(\bm{x}|\bm{y})\}.
\end{aligned}
\end{equation}
Note that due to the nonlinearity of the distribution system model, $\bm{f(\cdot)}$, an explicit expression of $\pi_{\text{post}}(\bm{x}|\bm{y})$ is extremely difficultly to derive. This is especially true for our topology estimation problem, where a group of $0$-$1$ binomial distributions  representing the status of the switches and another group of continuous random variables representing unknown system states are considered simultaneously. 
This motivates us to leverage the AIS method to recover all the Bayesian posterior distributions for the unknown parameters following different types of distributions. 

\vspace{-0.2cm}
\section{Adaptive Importance Sampling}
Although IS is more widely known in the realm of rare-event estimation \cite{chen2013composite,bugallo2017adaptive}, in this section, we will present the recovery of the  Bayesian posteriors using the weights of the IS. Then, we will further elaborate a more cost-effective AIS. 

\vspace{-0.2cm}
\subsection{Importance Sampling}
Let us present the principle of the importance sampling (IS) method. Following \cite{bugallo2017adaptive}, this method consists in drawing $K$ independent samples, ${\{\bm{x}^{(k)}\}}_{k=1}^{K}$,  samples from the proposal pdf, ${q}(\bm{x})$, which has heavier tails than the target function, ${\pi_{T}}(\bm{x})$, does. Each sample has an associated importance weight given by
\begin{equation}
    w^{(k)}=\frac{{\pi_{T}}(\bm{x}^{(k)})}{{q}(\bm{x}^{(k)})}, \;\;\;   k=1,\dots,K,
\end{equation}
where $w^{(k)}$ represents the importance of the sample $\bm{x}^{(k)}$ for the approximated target function ${\pi_{T}}(\bm{x})$ given the proposal function ${q}(\bm{x})$. 
For the parameter estimation problem we have considered in this paper, the target function, ${\pi_{T}}(\bm{x})$, comes from the Bayesian posterior pdf, $\pi_{\text{post}}(\bm{x}|\bm{y})$;
the pdf, ${q}(\bm{x})$, comes from some prior belief and typically has heavier tails than the posteriors. 
This enables us to draw a sample set from  ${q}(\bm{x})$ to represent parameter uncertainties as ${\{\bm{x}^{(k)}\}}_{k=1}^{K}$ and  evaluate the weight $w^{(k)}$ at each parameter value through   
\begin{equation}
\label{ISwiehgt}
    w^{(k)}=\frac {\pi_{\text{post}}(\bm{x}^{(k)}|\bm{y})}{{q}(\bm{x}^{(k)})}, \;\;\;   k=1,\dots,K.
\end{equation}
Then, the normalized weight $\bar{w}^{(k)}$ is calculated as
\begin{equation}
\label{NormISwiehgt}
    \bar{w}^{(k)}=\frac{w^{(k)}} {\sum_{i=1}^{K} w^{(i)}}.
\end{equation}
The normalized weights ${\{\bar{w}^{(k)}\}}_{k=1}^{K}$ allow us to recover the full probability distributions of $\pi_{\text{post}}(\bm{x}|\bm{y})$ via 
\begin{equation}
\label{recover}
    \pi_{\text{post}}(\bm{x}|\bm{y})=\sum_{k=1}^{K}\bar{w}^{(k)}\delta(\bm{x}-\bm{x}^{(k)}),
\end{equation}
where $\delta$ represents the Dirac delta function. {More specifically, to obtain the posterior distributions using the normalized weights, we suggest the use of equal-weight samples illustrated in detail in~\cite[\S2.2]{li2015adaptive}. Also, for the readers' implementation convenience, it is straightforward to use the built-in {\ttfamily randsample} function in MATLAB\textsuperscript{\textregistered} platform to achieve the recovering procedure.}

Using the obtained non-Gaussian posterior distribution for $\pi_{\text{post}}(\bm{x}|\bm{y})$, we can obtain the estimated model parameters via the aforementioned MAP estimator in (\ref{MAP}).
Also, the mean for $\pi_{\text{post}}(\bm{x}|\bm{y})$ can be further obtained via \cite{el2018robust}
\begin{equation}
\label{ismean}
    {\bm{\mu_{x}}}=\sum_{k=1}^{K}\bar{w}^{(k)}\bm{x}^{(k)}.
\end{equation}

The detailed steps for estimating the unknown parameters via IS have been summarized in Algorithm 1. It can be seen that, although IS is known as a variance-reduction technique to accelerate the Monte-Carlo sampling, Step 5 can still be time-consuming since the distribution system model $\bm{f(\cdot)}$ is repeatedly evaluated for the recovery of the Bayesian posteriors, (\ref{eq:post}). Besides, as shown in Step 1, $q(\bm{x})$ is obtained through an initial guess, which might be quite inaccurate in practice, diminishing the efficiency of the IS scheme. To overcome these shortcomings, we introduce a more advanced AIS scheme next.

 {We would like to emphasize that the importance sampling technique has been widely applied in power system rare-event simulations~\cite{chen2013composite}, secure operation design~\cite{de2018security}, reliability assessment~\cite{de2012simplified}, risk assessment~\cite{da2018risk}, to cite a few. It can also be flexibly combined with other techniques (e.g., cross-entropy~\cite{de2018security,da2018risk}~and antithetic variate~\cite{chen2013composite}) to further improve its estimation accuracy and computational efficiency. However, all of the above cited papers use the properties of IS for a forward uncertainty quantification problem. On the contrary, our work adopts the IS technique for a typical inverse uncertainty quantification problem, i.e., the topology estimation problem. A more detailed review of IS applications is provided in~\cite{bugallo2017adaptive}.
}

\begin{algorithm}[!htbp]
\label{IS}
\caption{Importance-Sampling-Based Bayesian Inference for Parameter Estimation}
\begin{algorithmic} [1]
\State  Set  proposal function $q(\bm{x})$ based on an initial guess; 
\State Construct the model $\bm{f(\cdot)}$, i.e., the distribution system model in Section II-A;
\State Draw a sample set, ${\{\bm{x}^{(k)}\}}_{k=1}^{K}$, from $q(\bm{x})$;
\For{$k=1,\ldots,K$}
\State Evaluate Bayesian posteriors's likelihood at sample values for   ${\pi_{\text{post}}(\bm{x}^{(k)}|\bm{y})}$ via (\ref{eq:post});
\State Evaluate weights for all samples via  (\ref{ISwiehgt});
\State Normalize the weights via (\ref{NormISwiehgt});
\EndFor
\State Recover the pdfs for $\pi_{\text{post}}(\bm{x}|\bm{y})$ via (\ref{recover});
\State Use MAP to estimate parameters via (\ref{MAP}).
\end{algorithmic}
\end{algorithm}

\vspace{-0.3cm}
\subsection{Adaptive Importance Sampling}
The AIS method is based on an iterative process for gradual evaluation of the proposal functions to accurately approximate the posterior functions~\cite{bugallo2017adaptive}. This AIS method consists of three basic steps: (i) generate samples from proposal functions; (ii) calculate weights for samples; and (iii) update the  parameters that define the proposals to obtain the new proposal for further iterations.

More specifically, for our parameter estimation problem with $N_{x}$ unknown parameters, the AIS algorithm is initialized with a set of $N_{x}$ proposals $\{q_{n}(\bm{x}|\bm{\Theta_{n,1}})\}_{n=1}^{N_{x}}$. {Each proposal is parameterized by a vector $\bm{\Theta_{n,1}}$}, which can initially come from the Bayesian prior pdfs, $\pi_{\text{prior}}(\bm{x})$. After drawing a set of samples $\bm{x}_{n,1}^{k},n=1,\dots,N_{x}, k=1,\dots,K$, we can obtain the normalized weights. These weights enable us to obtain a discrete probability distribution that approximates the target Bayesian posteriors $\pi_{\text{post}}(\bm{x})$ via (\ref{recover}) or the mean of the Bayesian posteriors via \eqref{ismean}. Then, the parameters of the $n$th proposal are updated from $\bm{\Theta}_{n,1}$ to $\bm{\Theta}_{n,2}$ based on the $n$th Bayesian posterior in $\pi_{\text{post}}(\bm{x})$. This process is repeated to make $\bm{\Theta}_{n,j}$ in the $j$th iteration to move to $\bm{\Theta}_{n,j+1}$ in the ($j+1$)th iteration until an iterative stopping criterion is satisfied.  Similarly, we  can obtain the estimated  parameters either via the  MAP estimator using \eqref{MAP} or via the mean estimator using \eqref{ismean} in the last iteration as the final results. This updating enables us to find a better proposal function that will allow us to draw more samples from the sample space with high likelihood and, therefore, will increase the estimation accuracy and efficiency of the AIS-based Bayesian inference.  Here, for the $0$-$1$ binomial distributions, parameterized by a success probability $p_\text{bino}\in[0,1]$, representing the status of the switches, if we obtain the MAP or the mean of the switch's posterior with a success probability greater than $0.5$, then we perceive it as a closed one, and vice versa \cite{zhao2019learning}.

\vspace{-0.3cm}
\section{The Proposed method}
In this section, we will formulate the detailed Bayesian inference model for the topology, outage, and states joint estimation with very few measurements using the AIS technique. 
\vspace{-0.6cm}
\subsection{ Bayesian Formulation of the Topology Estimation}
\subsubsection{Traditional Bayesian Formulation} First, let us present one possible Bayesian model for our topology estimation problem as 
\begin{equation}
\label{bayf1_short}
\bm{\hat{B}}_{\text{MAP}}=\underset{\bm{B}}{\text{arg min}}   \quad  \{-\pi_{\text{post}}(\bm{B}|\bm{y})\}.
\end{equation}
\normalsize
Starting with the measurement model, let us define the per-phase measured value of the active and reactive power in a distribution line  as  $\{{{P}_{L}^{\mathcal{M}}}^{\phi}\}_{{\phi}\in\mathcal{P}_{m, n}}$ 
and $\{{{Q}_{L}^{\mathcal{M}}}^{\phi}\}_{{\phi}\in\mathcal{P}_{m, n}}$; the per-phase forecast value of the active and reactive power in the bus as $\{{{P}_{F}^{\mathcal{M}}}^{\phi}\}_{{\phi}\in\mathcal{P}_{n}}$ 
and $\{{{Q}_{F}^{\mathcal{M}}}^{\phi}\}_{{\phi}\in\mathcal{P}_{n}}$; and the per-phase metered value of the active and reactive power in the end-user bus as $\{{{P}_{E}^{\mathcal{M}}}^{\phi}\}_{{\phi}\in\mathcal{P}_{n}}$ 
and $\{{{Q}_{E}^{\mathcal{M}}}^{\phi}\}_{{\phi}\in\mathcal{P}_{n}}$, respectively. They all may be modeled as measurement errors added to their true values, that is, 
${{P}_{L}^{\mathcal{M}}}={P}_{L}+e_{PL}$, 
${{Q}_{L}^{\mathcal{M}}}={Q}_{L}+e_{QL}$, ${{P}_{F}^{\mathcal{M}}}={P}_{F}+e_{PF}$,
${{Q}_{F}^{\mathcal{M}}}={Q}_{F}+e_{QF}$, 
${{P}_{E}^{\mathcal{M}}}={P}_{E}+e_{PE}$,
${{Q}_{E}^{\mathcal{M}}}={Q}_{E}+e_{QE}$.
Here, $e_{PL}$, $e_{QL}$, $e_{PF}$, $e_{QF}$, $e_{PE}$, and $e_{QE}$ are subvectors of the  measurement-error vector, $\bm{e}$, in \eqref{memodel}.
It is clear that a metered value has a smaller error compared to a pseudomeasurement, i.e., the forecast data. For example, we can assume that $e_{PL}$, $e_{QL}$, $e_{PE}$, and $e_{QE}$ are independent and identically distributed (i.i.d.) Gaussian error with a standard deviation of $0.1\%$ or $1\%$ while the values for $e_{PF}$ and $e_{QF}$ can range from $5\%$ to $15\%$. {Note that in practice, different types of loads (e.g., commercial, industrial, and residential) can exhibit different statistical properties. For instance, their standard deviation or ranges depend on the accuracy of both the measurement devices and the forecast procedure. Although we do not address this problem in this paper, these statistics should be carefully chosen for a better estimation performance. However, since our algorithm is a general Bayesian approach whose performance barely depends on them, it is rather  straightforward to adjust their values based on the implementation conditions. This will be further discussed in Section V-B. Also, the forecast errors, $e_{PF}$ and $e_{QF}$, of our method are assumed to be known. The development of a detailed forecasting technique goes beyond the scope of this paper.}

To be more realistic, let us assume that we only have one meter placed in the primary feeder to measure the power in the line, and a small portion (e.g., $15\%$ and $30\%$) of the end-users have measurement devices.  All the other end-users that have no meters rely on the forecast data seen as pseudomeasurements, which are much less accurate. {Note that the measured quantities can vary in practice. In our framework, although the quantities assumed to be metered are power measurements, they may equally be voltage or current measurements.} Within this framework and using the aforementioned distribution power-flow model, the Bayesian inference framework can be formulated as 
\begin{equation}
\label{bayf1}
\resizebox{.88\linewidth}{!}{ $\bm{\hat{B}}=\underset{\bm{B}}{\text{arg min}}   \quad  \{-\pi_{\text{post}}(\bm{B}|\{{{P}_{L}^{\mathcal{M}}}^{\phi},
{{Q}_{L}^{\mathcal{M}}}^{\phi},
{{P}_{F}^{\mathcal{M}}}^{\phi},
{{Q}_{F}^{\mathcal{M}}}^{\phi},
{{P}_{E}^{\mathcal{M}}}^{\phi},
{{Q}_{E}^{\mathcal{M}}}^{\phi}
\})\}$ }.
\end{equation}
The only optimized variables in this formulation is the vector of the binary variables, $\bm{B}$, that can account for the topology uncertainties brought by switch statuses \cite{cavraro2019real,gandluru2019joint}.  

However, in practice, the power injections from the end-users are also unknown. They represent the unknown states in the system that can also influence the model output in the distribution system. Therefore, an alternative formulation is proposed as follows:
\begin{equation}
\label{bayf2}
\resizebox{.88\linewidth}{!}{ $\underset{\bm{B},\bm{P}_{b},\bm{Q}_{b}}{\text{min}}   \quad  \{-\pi_{\text{post}}(\bm{B},\bm{P}_{b},\bm{Q}_{b}|
\{{{P}_{L}^{\mathcal{M}}}^{\phi},
{{Q}_{L}^{\mathcal{M}}}^{\phi},
{{P}_{F}^{\mathcal{M}}}^{\phi},
{{Q}_{F}^{\mathcal{M}}}^{\phi},
{{P}_{E}^{\mathcal{M}}}^{\phi},
{{Q}_{E}^{\mathcal{M}}}^{\phi}
\})\}$ }.
\end{equation}
In this new formulation, the optimization problem becomes more complicated since both the switch statuses and state variables are estimated jointly. 

\subsubsection{Drawbacks of Using Pseudomeasurements}  As we state in Section I, due to the radial structure of the distribution system, the topology change in distribution system is more likely to induce outages. Therefore, the operation topology estimation should account for the outage estimation as well~\cite{gandluru2019joint}. 

However, within the outage area, where the 
the distribution lines and loads are not energized, the forecast data cannot act like a ``real measurement'' to reflect the de-energized load, but remain unchanged and, therefore, become  outliers that can fully bias the estimator formulated in \eqref{bayf1} or \eqref{bayf2}. Currently, one solution in the literature is to seek help from the ping measurement that can check the connectivity of the load to ensure the proper usage of pseudomeasurement, which allows us to still use \eqref{bayf1} or \eqref{bayf2}.  However, to the best of our knowledge, the ping measurements have not yet been widely deployed in practice due to issues related to privacy, cost, etc.  Therefore, we choose not to rely on the ping measurement in this paper, but rather we propose a two-stage estimation procedure as proposed next.

\subsubsection{Bayesian Reformulation Facing Outages} Within the two-stage estimation procedure, this first-stage procedure remains the Bayesian inference. But, we propose to formulate it as
\begin{subequations}
\label{bafinal}
\begin{align}
\underset{\bm{B},\bm{P}_{b},\bm{Q}_{b}}{\text{min}}   \quad  & \{-\pi_{\text{post}}(\bm{B},\bm{P}_{b},\bm{Q}_{b}|\{{{{P}_{L}^{\mathcal{M}}}^{\phi},
{{Q}_{L}^{\mathcal{M}}}^{\phi},
{{P}_{E}^{\mathcal{M}}}^{\phi},
{{Q}_{E}^{\mathcal{M}}}^{\phi}} \})\} \label{eq:subf}\\
\textrm{s.t.} \quad & \mathbf{g}(\bm{B},\bm{P}_{b},\bm{Q}_{b})=0\\
  & {\bm{P}_{b}}^{l} \leq \bm{P}_{b} \leq {\bm{P}_{b}}^{u} \label{eq:subp}\\
   & {\bm{Q}_{b}}^{l} \leq \bm{Q}_{b} \leq {\bm{Q}_{b}}^{u} \label{eq:subq}.
\end{align}
\end{subequations}
As is shown in \eqref{eq:subf}, the reformulation eliminates the pseudomeasurement from the measurement model to avoid to use of biased data in the outage area. 
The equality constraints from $\mathbf{g}(\boldsymbol{\cdot})$ represent the physics-constrained power-flow equations given the parameters, $\{\bm{B},\bm{P}_{b},\bm{Q}_{b}\}$. 
In order not to waste the information in the forecast data, we introduce the lower bounds, ${\bm{P}_{b}}^{l}$ and ${\bm{Q}_{b}}^{l}$, and upper bounds, ${\bm{P}_{b}}^{u}$ and ${\bm{Q}_{b}}^{u}$, for the optimized states, $\bm{P}_{b}$ and $\bm{Q}_{b}$. For the pseudo-metered buses at the user-ends,
these bounds are calculated from the forecast data. Let us take the mean and the standard deviation for the $\bm{P}_{b}$ and $\bm{Q}_{b}$ as the $\bar{\bm{P}}_{b}$, $\bar{\bm{Q}}_{b}$,  $\delta_{p_{b}}$ and the $\delta_{q_{b}}$. By taking the forecast data ${{P}_{F}^{\mathcal{M}}}^{\phi}$ and ${{Q}_{F}^{\mathcal{M}}}^{\phi}$ as the  values for $\bar{\bm{P}}_{b}$ and $\bar{\bm{Q}}_{b}$, we can set the bounds as
${\bm{P}_{b}}^{l}=\bm{P}_{b}-3\delta_{p_{b}}$, ${\bm{P}_{b}}^{u}=\bm{P}_{b}+3\delta_{p_{b}}$, ${\bm{Q}_{b}}^{l}=\bm{Q}_{b}-3\delta_{q_{b}}$, and ${\bm{Q}_{b}}^{u}=\bm{Q}_{b}+3\delta_{q_{b}}$ since it can cover the $99.7\%$ probability under a {Gaussian assumption}\footnote{{Here, it is worth emphasizing that although we adopt the Gaussian assumption like most of the existing literature on topology estimation, the behaviors of the load in practical power distribution systems can follow a different distribution and sometimes exhibit discrete jumps. Accordingly, we need to adjust the pdfs to improve the modeling accuracy. For example, for the load modeled with some discrete distributions, the discrete Poisson distribution may serve as a suitable candidate. Furthermore, for some loads that exhibit more complex behaviors, a hybrid technique may be  considered~\cite{mele2019modeling}.  In this paper, we solely apply the Gaussian assumption for simplicity; the detailed load modeling issue and its associated forecasting techniques are outside the scope of this paper.}} for the forecast error. The same logic applies to the power injections at the metered buses using the metered data ${{P}_{E}^{\mathcal{M}}}^{\phi}$ and ${{Q}_{E}^{\mathcal{M}}}^{\phi}$.
By this way, the reformulation in \eqref{bafinal} can not only make use of pseudomeasurements, but also avoids the bias induced by errors in outage estimation.

\subsubsection{An Additional Island-Component Detection Procedure}
Although the above reformulating in~\eqref{eq:subf} can avoid the biased estimation results in the non-outage area (i.e., the energized region), it cannot avoid the biased results in the outage region where no meter is placed. Therefore, the second-stage estimation will focus on the correction in the outage area. Following the first procedure, we can obtain the estimated statuses of the switches, $\bm{\hat{B}}$, from which we can subsequently recover the structure of the distribution grid. With this structure, we can identify the components located in the outage area as \emph{island components}. Then, we set the estimated $\hat{\bm{P}}_{b}$ and $\hat{\bm{Q}}_{b}$ in this area to be $0$, and identify the switches in this area as the \emph{inestimable} ones. Till now, we have completed the presentation of the two-stage estimation procedure that enables the topology, outage, and state joint estimation.

\subsection{AIS-enhanced Bayesian Inference}
Using the above two-stage estimation procedure and the AIS technique, the detailed procedure for the proposed method is described in Algorithm 2.

\begin{algorithm}[!htbp]
\label{MF-PCE-AIS}
\caption{A Bayesian Approach for Distribution System Topology Estimation via AIS}

\begin{algorithmic} [1]
\State  Set proposal functions $\{q_{n}(\bm{x}|\bm{\Theta_{n,1}})\}_{n=1}^{N_{x}}$ for the parameters,  $\{\bm{B},\bm{P}_{b},\bm{Q}_{b}\}$, and initiate $\{\bm{\Theta}_{n,0}\}_{n=1}^{N_{x}}$;
\State  Formulate the distribution system model and its measurement model with $\{-\pi_{\text{post}}(\bm{B},\bm{P}_{b},\bm{Q}_{b}|\{{{{P}_{L}^{\mathcal{M}}}^{\phi},
{{Q}_{L}^{\mathcal{M}}}^{\phi},
{{P}_{E}^{\mathcal{M}}}^{\phi},
{{Q}_{E}^{\mathcal{M}}}^{\phi}} \})\}$;
 \State Initiate iteration number $j$;

\While{{(stopping criterion is not met)} }

\For {$k=1,\ldots,K$}

\State Draw the proposed sample set, ${\{\bm{B}^{(k)}, {\bm{P}_{b}}^{(k)},{\bm{Q}_{b}}^{(k)}\}}_{k=1}^{K}$, from $\{q_{n}(\bm{x}|\bm{\Theta_{n,1}})\}_{n=1}^{N_{x}}$;

\State Evaluate the Bayesian posterior likelihood at sample values for  
\small
\footnotesize
$\{-\pi_{\text{post}}(\bm{B}^{(k)}, {\bm{P}_{b}}^{(k)},{\bm{Q}_{b}}^{(k)}|\{{{{P}_{L}^{\mathcal{M}}}^{\phi},
{{Q}_{L}^{\mathcal{M}}}^{\phi},
{{P}_{E}^{\mathcal{M}}}^{\phi},
{{Q}_{E}^{\mathcal{M}}}^{\phi}}\})\}$
\normalsize
via  \eqref{eq:subf};


\State Evaluate weights for all samples via  (\ref{ISwiehgt});

\State Normalize the weights via (\ref{NormISwiehgt});

\State Use MAP or mean estimator to approximate parameters via (\ref{MAP}) or \eqref{ismean};

\State Update $\{\bm{\Theta}_{n,j}\}_{n=1}^{N_{x}}$  to $\{\bm{\Theta}_{n,j+1}\}_{n=1}^{N_{x}}$ from $\pi_{\text{post}}(\bm{x}|\bm{y})$ to get new  proposal functions $\{q_{n}(\bm{x}|\bm{\Theta_{n,j+1}})\}_{n=1}^{N_{x}}$;

\EndFor
\State Update $j=j+1$;
\EndWhile
\State Read the estimation results for $\{\hat{\bm{B}}, \hat{\bm{P}}_{b}, \hat{\bm{Q}}_{b}\}$;

\State Recover the structure of the grid via switch statuses, $\hat{\bm{B}}$;

\State Correct the estimated variables in the outage area. 

\end{algorithmic}
\end{algorithm}

In Algorithm 2, the stopping criterion is obtained by setting a threshold, $j_{\max}$, to the maximum number of iterations (e.g., $4$ or $6$) that can be tuned accordingly. Moreover, to update the parameters of the proposal function from $\{\bm{\Theta}_{n,j}\}_{n=1}^{N_{x}}$ to $\{\bm{\Theta}_{n,j+1}\}_{n=1}^{N_{x}}$, there exist different strategies, such as population Monte Carlo (PMC) \cite{cappe2004population}, deterministic-mixture PMC \cite{elvira2017improving}, and adaptive multiple IS \cite{cornuet2012adaptive}. Here, we choose the PMC for its simplicity. In this scheme, we only need to update the location parameters of the proposal functions for the next iteration~\cite{bugallo2017adaptive}. These location parameters can be easily obtained from the MAP or the mean estimator for the recovered Bayesian posteriors at the current iteration. For the continuous variables, $\bm{P}_{b},\bm{Q}_{b}$, if a sample drawn from the proposal function goes beyond the bounds as listed in \eqref{eq:subp} and \eqref{eq:subq}, then we can simply place them at the value of the bound.  Since it is suggested to have heavy tails for the proposal function in \cite{bugallo2017adaptive}, $p_\text{bino}$ can be extremely close to $0$ or $1$ for some scenarios. To  maintain a relatively thicker tails, we can  simply set the lower and upper bounds for the value of $p_\text{bino}$ in the proposal functions, to $0.15$ and $0.85$, respectively. Now, we have completed the presentation of parameter tuning in the AIS.  

We would also like to emphasize that although our framework can simultaneously approximate the topology, outage, and states, our initial and major goals are the topology and outage joint estimation. It should be noted that the state estimation comes as a byproduct of our estimator. Furthermore, since it is well-known that the number of the possible topologies can be  approximated as  $2^{N_s}$, which requires an exhaustive search as ${N_s}$ grows large, we do not expect the AIS method to always approach the global optimal for $\hat{\bm{B}}$ due to the nonlinearity of the model, the scale and NP-hardness of the problem, and limited measurements. Yet, we are still able to use a MAP or mean estimator to obtain the switch status by judging the success probability, $p_\text{bino}$, for the binomial distribution of a switch. Note that, if we get a value of $p_\text{bino}$ very close to $0.5$ (e.g., $0.45$ to $0.55$) for a switch, that means the Bayesian posterior reflects a solution with a low confidence; then, we suggest the distribution system operator not to trust the current estimation result for such a switch. Till now, we have completed the presentation of the proposed AIS-enhanced Bayesian approach in the topology, outage, and state joint estimation for the unbalanced distribution system.

\vspace{-0.2cm}
\section{Simulation Results}
\begin{figure}[!htbp]
 \centering
 \includegraphics[scale=0.55]{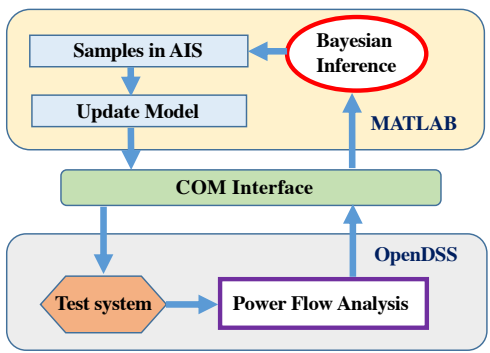}
 \setlength{\abovecaptionskip}{-2pt}
 \caption{MATLAB\textsuperscript{\textregistered}-OpenDSS co-simulation environment for the implementation framework.}
 \label{fig:1}
 \end{figure}

\vspace{-0.1cm}
Using the proposed method, various case studies are conducted on the modified IEEE 123-bus test system and a modified unbalanced  1282-bus system. Their data can be accessed through the Open Distribution System Simulator (OpenDSS) package \cite{dugan2011open}.
The simulation framework is tested with the MATLAB\textsuperscript{\textregistered} R2020b version on a laptop with 2.60-GHz Intel\textsuperscript{\textregistered} Core\texttrademark~i7-6600U processors and a 16 GB of main memory. The unbalanced distribution system is modeled and calculated in the OpenDSS. More specifically, we use MATLAB\textsuperscript{\textregistered} to control the OpenDSS through a component object model (COM) interface that allows us to change the parameters for loads, closed/open  switches, and evaluate power-flow solutions \cite{theodoro2018matlab}.  {Here, as suggested in \cite{gandluru2019joint} and \cite{ardakanian2019identification}, the control mode is disabled in the OpenDSS in order to ensure that the transformer taps are not automatically adjusted. Thus, we can focus on the switch-status-induced topology, outage, and state joint estimation problem. The parameters of the  model components are assumed to be known based on our discussion in Remark 1.} The general framework for implementing our proposed method in the MATLAB\textsuperscript{\textregistered}-OpenDSS co-simulation environment is depicted in Fig.~\ref{fig:1}. Then, various case studies are conducted to validate the performances of the proposed method. 

\vspace{-0.3cm}
\subsection{Demonstration on the IEEE 123-bus Test System}
\subsubsection{Experiment Settings}
\begin{figure}[!htbp]
\vspace{-0.22cm}
 \centering
 \includegraphics[scale=0.36]{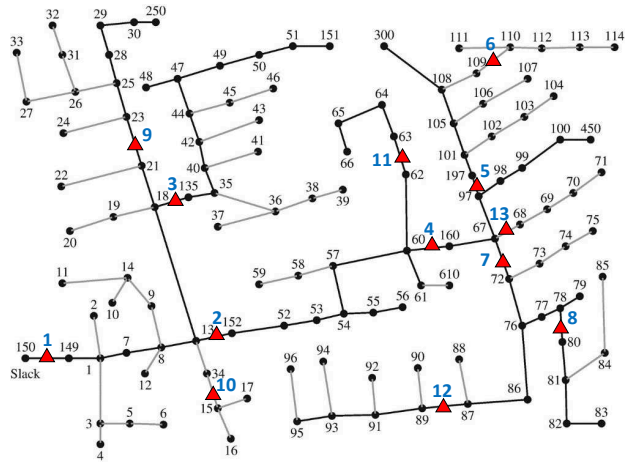}
 \setlength{\abovecaptionskip}{-2pt}
 \caption{Topology of the IEEE 123-bus system.}
 \label{fig:2}
 \end{figure}

First, let us present a demonstration of the proposed method applied in a small-scale IEEE 123-bus system, which is well-known for its unbalanced structure that consists of $3$-, $2$-, and $1$-phase, distribution lines associated with $91$ loads with different types of  connections. Its topology and the location of the $13$ switches are shown in Fig. \ref{fig:2}. 
Here, let us assume that $30\%$ of the end-users are equipped with the meters, whose measurement errors,
 $e_{PE}$ and $e_{QE}$, is set to have an i.i.d. Gaussian distribution with a standard deviation of  $1\%$, while the remaining $70\%$ of the end-users are  using forecast data with their errors, $e_{PF}$, $e_{QF}$,  following the i.i.d.  Gaussian assumption, whose standard deviation is $10\%$ with respect to their mean. To explore the status of the $13$ switches in the test system, an exhaustive exploration will require $2^{13}=8,192$ tests for all possible topologies. This task is nontrivial since we also need to estimate the unknown system states. Here, let us first use this test system to demonstrate the efficiency of the AIS method in the topology estimation by using a sample size much smaller than $8,192$ while providing an accurate estimation result.  To make the estimation task more challenging, we set the true state vector of the switches to be $\bm{B}=[1,     1,     0,     1,     0,     0,     1,     0,     0,     1,     1,     0,     1]$. In this case, nearly half of the switches are open and, therefore, inducing multiple outage areas simultaneously. Note that in the second stage for the correction elaborated in Section IV-A4, use of \eqref{eq:post} is quite necessary. For example, since Switch $5$ is open, Switch $6$ is de-energized and, therefore, needs to be identified as the inestimable switch. To make a fair comparison, we  conduct the estimation $100$ times separately to calculate {the estimation accuracy}\footnote{{Here, the estimation accuracy is defined as the ratio of the switch status estimated correctly. For example, for $10$ switches, if $9$ of their statuses are correctly estimated, then the accuracy is said to be $90\%$.}} for all the switch  statuses using the AIS with different maximum iteration number, $j_{\max}$.
\vspace{-0.2cm}
 \begin{table}[hbtp]
\renewcommand{\arraystretch}{1.2}
\caption{Accuracy of the Proposed Estimation Method Applied to the Modified IEEE 123-bus System}
\label{ieee123}
\centering
\begin{tabular}{cccc}
\toprule
{} & $j_{\max}=1$ & $j_{\max}=2$  & $j_{\max}=3$  \\
\hline
$\rho_{1{\text{st}}} [\%]$ & $86.15$ & $96.15$ & $96.62$   \\
$\rho_{2{\text{nd}}} [\%]$ & $88.77$ &  $99.54$  & $100$  \\
Time [s] & $8.9$ & $ 17.1$  & $24.3$  \\
\bottomrule
\end{tabular}
\vspace{-0.4cm}
\end{table}
\subsubsection{Comparison studies using different iterations}
From Table~\ref{ieee123}, it can be seen that with only ordinary IS without iteration (i.e., $j_{\max}=1$), the proposed method can still correctly approximate most statuses of the switches. More specifically, approximately $12$ out of $13$ switches are accurately estimated with the MAP while only less than $1/8$ of all the possible topologies have been explored. Further, once we use the AIS, even the iteration number is low (e.g., $j_{\max}=2$), it becomes almost impossible to obtain an incorrect estimation for a switch position although the estimation time increases. Here, thanks to the super-fast calculation speed of the OpenDSS, thousands of samples can still be evaluated in a reasonably short time, rendering it applicable for online applications.   
 
\subsubsection{Capability of State Estimation}
\begin{figure}[!htbp]
 \centering
 \includegraphics[scale=0.27]{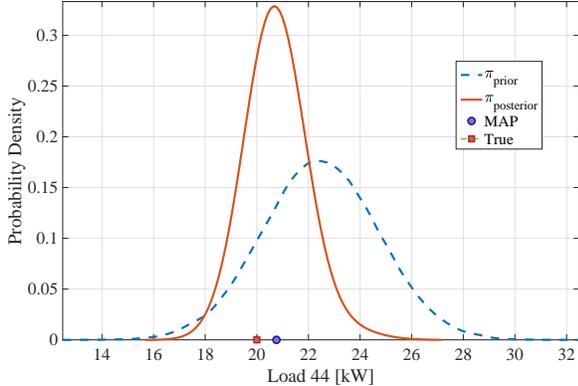}
 \setlength{\abovecaptionskip}{-5pt}
 \caption{Bayesian  posterior for the active power of Load 44. The MAP estimate (\textcolor{blue}{blue} circle) is at $20.6522$ kW while the true value ({red} square) is at $20$ kW.}
 \label{fig:3}
 \end{figure}

It is also worth pointing out that  {our Bayesian formulation has the natural ability to estimate the power injections.}\footnote{{Here, we specifically mean that the load behavior is properly modeled, e.g., with a Gaussian distribution, whereas the aforementioned complicated load behaviors we discussed in Section IV-A are not considered.}} This is demonstrated in  Fig.~\ref{fig:3} by considering Load $44$. Here, we can get a Bayesian posterior distribution for this system state. However, we acknowledge that although the computing time of our algorithm is acceptable for the topology and outage estimation, it is not fast enough for a regular online state estimator. Therefore, we view the state estimation capability of our method as a byproduct\textemdash not the main contribution. 
 
 \subsubsection{{Key Parameter Tuning}}
{Like most of the Bayesian statistical inference algorithms (e.g., the Metropolis-Hastings mechanism \cite{xu2019response}), the parameter tuning needs to be performed carefully to have a good performance achieved by the IS-based Bayesian inference scheme. To demonstrate that, let us test the performances of our proposed method using different values for the bounds set for $p_\text{bino}$, which is briefly mentioned in Section IV-B. We set $j_{\max}=2$ since it has been proven to be a reliable value in the aforementioned test. The other settings remain unchanged. It is shown in Table~\ref{bounds} that the upper bounds range from $0.8$-$0.95$ (i.e., the lower bounds range from $0.05$-$0.2$), the accuracy is still quite high. However, when it comes to values of $0.99$, which indicates almost no bound is set since it is close to a probability of $1$, we can see a relatively larger drop in the estimation accuracy. This justifies setting such a bound. This is important in Bayesian inference since for a nonlinear optimization problem such as the one formulated in \eqref{bafinal}, setting a bound gives the algorithm a certain possibility to jump out from a local optimum to better search for a global optimum. Otherwise, the estimation accuracy will be inevitably reduced to some degree.}

\begin{table}[hbtp]
\centering 
\renewcommand{\arraystretch}{1.2}
\caption{Tests using Different Bounds for $p_\text{bino}$}
\label{bounds}
\begin{tabular}{ccccc}
\toprule
{$p_\text{bino}$} & $0.8$ & $0.85$  & $0.9$& $0.99$\\
\hline
$\rho_{2{\text{nd}}} [\%]$ & $99.69$ &  $99.54$  & $99.1$   & $96.23$\\
\bottomrule
\end{tabular}
\color{black}
\end{table}


\subsubsection{{Tests using Loop-Structured Distribution Network}} 
{In principle, the Bayesian method has no restriction on the structure of the system\textemdash be it radial-type or loop-type. The latter has been investigated by Zhao \emph{et al.} \cite{zhao2019learning} in the power transmission system topology identification problem, where the system structure is typically meshed and, therefore, \emph{non-radial}.}

\begin{figure}[!htbp]
\vspace{-0.22cm}
 \centering 
 \includegraphics[scale=0.336]{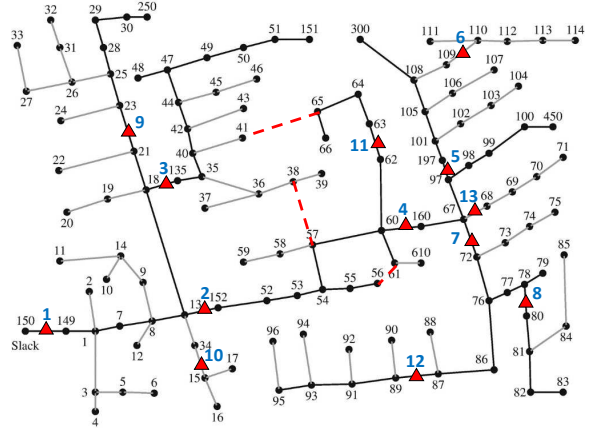}
 \setlength{\abovecaptionskip}{-2pt}
 \vspace{-0.3cm}
 \caption{Structure 1 of the modified topology of the IEEE 123-bus system.}
 \label{fig:loop}
 \end{figure}

\begin{figure}[!htbp]
\vspace{-0.22cm}
 \centering 
 \includegraphics[scale=0.336]{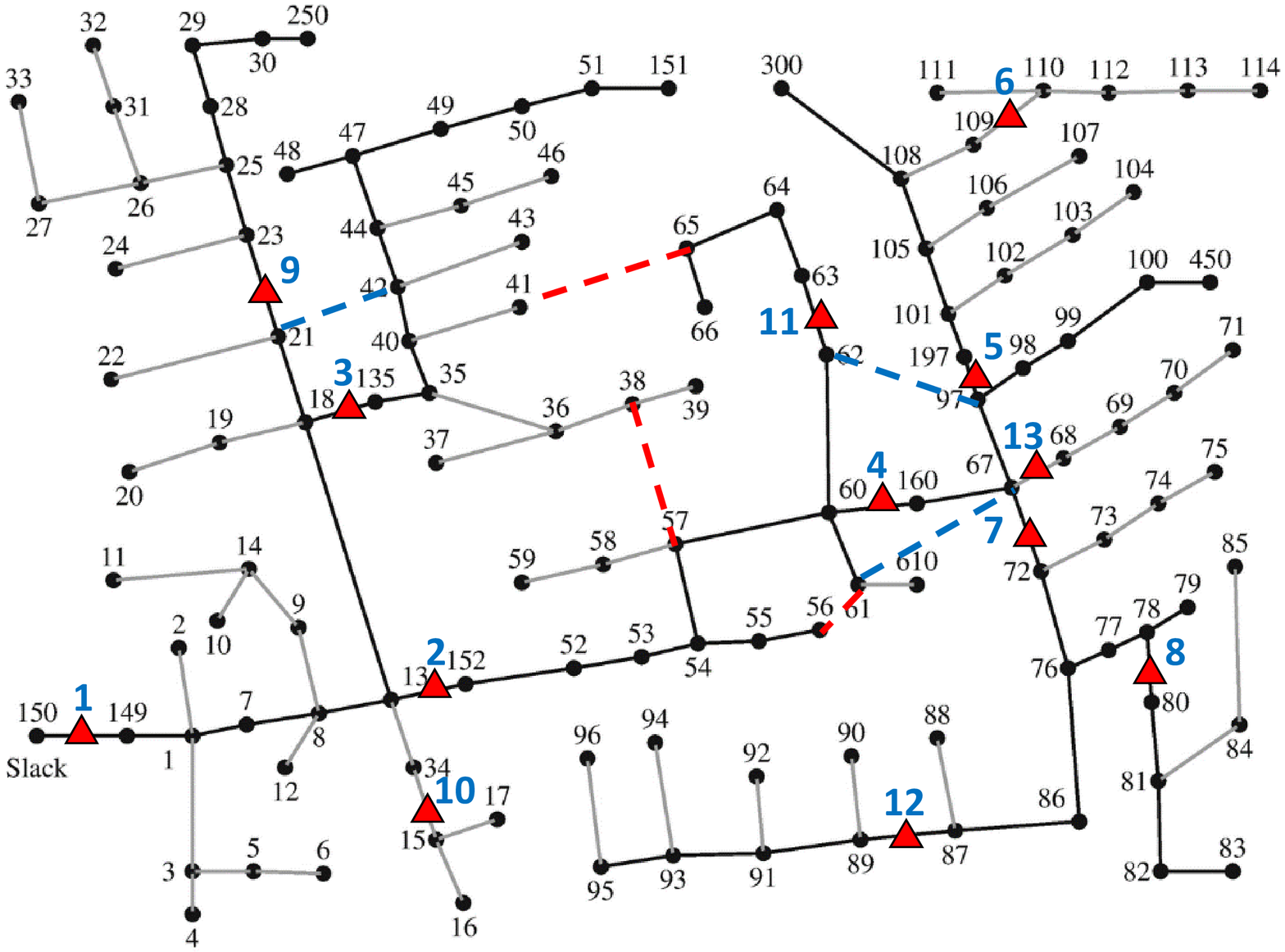}
 \setlength{\abovecaptionskip}{-2pt}
 \vspace{-0.3cm}
 \caption{Structure 2 of the modified topology of the IEEE 123-bus system.}
 \label{fig:moreloop}
 \end{figure}

{Since a loop structure may exist in urban power distribution systems, we further investigate the applicability of the proposed method to such systems. To this end, we modify the IEEE 123-bus system as shown in Fig.~\ref{fig:loop}, for which the dashed red lines are added  to create a loop structure of the network. The line  connecting Buses 56 and 61 is three-phase while the left two are single-phase. Then, we further increase the number of loops by adding three more three-phase lines as shown by the dashed blue ones in Fig.~\ref{fig:moreloop}. Again, we repeat the simulations conducted in Section V-A2 with these modified structures.  The simulation results are summarized in Table~\ref{ieee123loop}. It can be seen that the proposed method can provide quite accurate estimation results for both loop  structures. Besides, although the accuracy slightly drops when more loops are added, the AIS algorithm can still improve its accuracy by simply adding more iterations to fine-tune its results.  Therefore, we conclude that the Bayesian scheme does have the flexibility to perform well in a system with a loop structure, which is, indeed, an advantage compared to the spanning-tree algorithm, which assumes a radial structure.}
 
 \begin{table}[hbtp]
\renewcommand{\arraystretch}{1.2}
\centering 
\caption{Validation of the Proposed Estimation Method in the Modified IEEE 123-Bus System with Loop Structures}
\label{ieee123loop}
\begin{tabular}{ccccc}
\toprule
{} & {} &$j_{\max}=1$ & $j_{\max}=2$  & $j_{\max}=3$  \\
\hline
{Structure 1} & $\rho_{2{\text{nd}}} [\%]$ & $87.95$ &  $99.66$  & $100$  \\
{Structure 2} & $\rho_{2{\text{nd}}} [\%]$ & $88.6$ &  $96.4$  & $98.1$  \\
\bottomrule
\end{tabular}
\vspace{-0.4cm}
\end{table}

\subsection{Case Studies on a Larger-scale Test System}
\subsubsection{Validation of the Proposed Method}
Now, let us further validate the proposed method on a larger-scale test system, i.e., the unbalanced 1282-bus system located in the southeastern U.S {with
its 13-kV feeder supplying power to approximately 500 commercial and residential customers.} 
 Here, we  place $20$ switches in the system that can create $2^{20}=1,048,576$ possible topologies. For an online distribution system application, it is obviously not practical to have an exhaustive search over more than $1$ million possible topologies considering the computing time and the storage burden of the computing units. Again, in order to create a very challenging case, we set $8$ switches to be opened when the intent is to create multiple outages simultaneously. To make a fair comparison, we still conduct the estimation $100$ times separately to calculate the average estimate accuracy for different experiment settings. The detailed simulation results and the settings are provided in Table~\ref{ieee1282}.
\begin{table}[htbp]
  \centering
  \caption{Accuracy of the Proposed  Method under Different Estimation Conditions Applied to a 1282-bus system}
  \label{ieee1282}
  \begin{tabularx}{\linewidth}{*{13}{p{.10\linewidth}}}
  \toprule
    \multicolumn{7}{c}{ \textbf{Group $1$ under Different Measurement Accuracy}} \\ \toprule
{Samples}   &  {Iterations}  &   {Meter\newline Std. Dev.}  & Forecast\newline Std. Dev. &  Meter\newline Ratio  & {Time \newline [s]} & $\rho_{2{\text{nd}}}$ \newline $[\%]$  
\\ \hline 
  $1,000$  &  $6$ & $1\%$ &  $5\%$  &  $30\%$   &  $111$  &  $95.2$ \\
  $1,000$  &  $6$ & $0.5\%$ &  $5\%$  &  $30\%$   &  $112$  &  $95.86$ \\
 $1,000$  &  $6$ & $0.1\%$ &  $5\%$  &  $30\%$   &  $110$  &  $97.71$ \\
\bottomrule
  \end{tabularx}

  \begin{tabularx}{\linewidth}{*{13}{p{.10\linewidth}}}
  \toprule
    \multicolumn{7}{c}{ \textbf{Group $2$ under Different Forecast Accuracy}} \\ 
 \hline 
  $1,000$  &  $6$ & $1\%$ &  $5\%$  &  $30\%$   &  $111$  &  $95.2$ \\
  $1,000$  &  $6$ & $1\%$ &  $10\%$  &  $30\%$   &  $113$  &  $94.2$ \\
\bottomrule
  \end{tabularx}  

 \begin{tabularx}{\linewidth}{*{13}{p{.10\linewidth}}}
  \toprule
    \multicolumn{7}{c}{ \textbf{Group $3$ under Different Observability}} \\ 
 \hline 
  $1,000$  &  $6$ & $1\%$ &  $5\%$  &  $30\%$   &  $111$  &  $95.2$ \\
 $1,000$  &  $6$ & $1\%$ &  $5\%$  &  $20\%$   &  $110$  &  $92.57$ \\
 $1,000$  &  $6$ & $1\%$ &  $5\%$  &  $10\%$   &  $113$  &  $87$ \\
\bottomrule
  \end{tabularx}

 \begin{tabularx}{\linewidth}{*{13}{p{.10\linewidth}}}
  \toprule
    \multicolumn{7}{c}{ \textbf{Group $4$ for Comparison with IS}} \\ 
 \hline 
 $1,000$  &  $10$ & $1\%$ &  $10\%$  &  $30\%$   &  $203$  &  $94.64$ \\
 $1,000$  &  $10$ & $1\%$ &  $5\%$  &  $30\%$   &  $202$  &  $96.21$ \\
 $10,000$  &  $1$ & $1\%$ &  $10\%$  &  $30\%$   &  $202$  &  $84.2$ \\
  $10,000$  &  $1$ & $1\%$ &  $5\%$  &  $30\%$   &  $199$  &  $82.3$ \\
\bottomrule
  \end{tabularx} 
\end{table}

\begin{figure}[!htbp]
 \centering
 \includegraphics[scale=0.3]{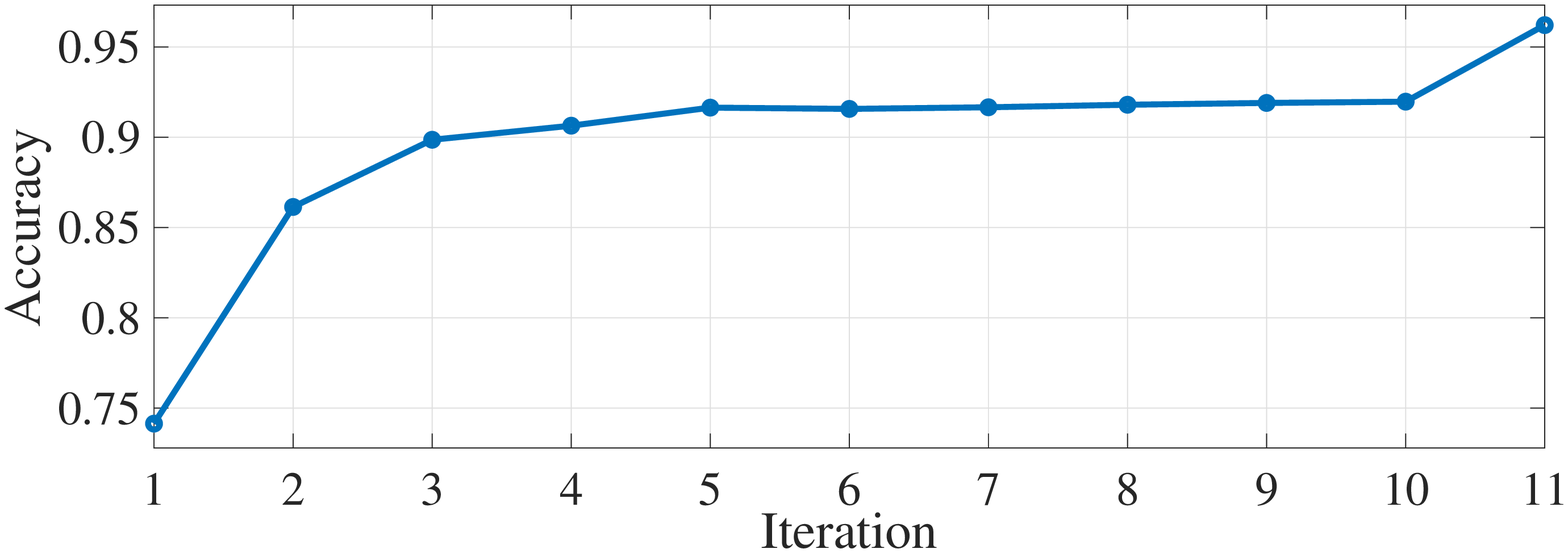}
 \setlength{\abovecaptionskip}{-5pt}
 \caption{Convergence plot of the proposed method.}
 \label{fig:4}
 \end{figure}

From Table~\ref{ieee1282}, the following conclusions can be drawn:

\begin{itemize}
  \item From the cases studies in Group $1$, we can see that the proposed method provides accurate estimation result under different levels. In general, with a smaller noise, the estimation accuracy increases slightly.  
  
  \item From the experiments in Group $2$, we can see that although the standard deviation of the errors in the forecast data have an impact on the estimation accuracy, the proposed method still provide a stable estimation results under a relatively large forecast error. 
  
  \item  From the observability tests in Group $3$, we can see that the ratio of the end-users that are equipped with the meters has a major impact on the estimation accuracy. In general, for this large system, to obtain a good estimation result,  the ratio should not be too low.

  \item In Group $4$, we conducted comparison studies between the AIS method and the IS method with the same amount of the total samples. It is quite clear that the incorporating of the adaptive procedure enables a better performance of the AIS method compared with the traditional IS method. Furthermore, we observe that by combining the cases in Groups $2$ and $4$, the number of iterations further increases, which only brings a marginal improvement in accuracy while significantly increasing the computing time. This is also observed in the estimation accuracy versus the number of iterations displayed in Fig.~\ref{fig:4}. Indeed, it can be seen that after approximately $5$ iterations, the estimation accuracy tends to level off. Note that the jump in the $11$th iteration is induced by the execution of the second-stage correction procedure.

  \item  In general, the proposed AIS method achieves a good estimation accuracy (i.e., around $95\%$), which means we can correctly estimate $19$ switches out of the $20$. The estimation accuracy still has the potential for further improvement if more end-users are equipped with meters with higher measuring accuracy or the forecast accuracy can be further improved. {We need to emphasize that the purpose of introducing the AIS method is to avoid the computing challenges met by the traditional exhaustive-search-based method, considering  the aforementioned 
 $2^{20}=1,048,576$ possible topologies, our AIS algorithm achieves a quite good estimation accuracy by only exploring a few thousand possible topologies, which is even less than $1\%$ of all the possible ones. This demonstrates a significant improvement compared with the exhaustive-search-based method. Finally, its computing time is typically less than $2$ min, which is acceptable for the topology and outage estimation in practice.}\footnote{{Here, as one reviewer has pointed out that although the computing speed of OpenDSS is fast, the communication between the simulation and inference blocks might be a performance bottleneck. This is, indeed, an important issue that needs to be addressed. In OpenDSS, there are two popular ways to import OpenDSS simulated data into the MATLAB\textsuperscript{\textregistered} platform for further inference. One is to first save the OpenDSS data into a .csv (or .txt) file. Then, we load the data of the Excel file into the MATLAB\textsuperscript{\textregistered} workspace. Another way is to directly read data from the COM Interface as shown in Fig.~\ref{fig:1}. In this way, we need to first set the active elements of OpenDSS simulator in the MATLAB\textsuperscript{\textregistered} platform (e.g., \emph{DSSObj.ActiveCircuit.Loads} for the loads, \emph{DSSObj.ActiveCircuit.Lines} for the network lines, etc.). Then, we can directly read their data from the COM Interface via MATLAB\textsuperscript{\textregistered}. Here, we found that the computational speed for the second way is much faster than the first way. Therefore, to maintain a high computational efficiency for the online application, it is very important to directly read the data from the COM Interface instead of using the Excel file. In this way, the communication challenge between the simulation and inference blocks can be greatly overcome to guarantee its computing efficiency for the online application.
  }}
  
  \end{itemize}

\subsubsection{{Validation of the Proposed Method under Different R/X Ratios}}
{  It is well known that a high R/X ratio in a power system can increase the nonlinearity of the model
  \cite{rajicic1988modification}, and sometimes can even lead to convergence issues of a power-flow solver \cite{mili2021alleviating}. This fact  holds especially true in power distribution systems, which typically exhibit a higher R/X than transmission systems do. Therefore, this incentivizes us to validate the performance of the proposed method by further increasing the R/X ratio in the original model~\cite{gandluru2019joint}.  To this end, we multiply the R/X ratio of the distribution lines of the original base case by different factors, e.g., $1.5$ and $2$. Again, we conduct the estimation $100$ times separately to calculate the averaged estimation accuracy.
  The simulation results are shown in Table~\ref{RXratio}. It is demonstrated that even when the nonlinearity of the system model is increased, we still obtain reasonably good estimation results after only $6$ iterations and using a meter ratio of $30\%$. Here, the accuracy of the results only drops slightly with an increase in the R/X ratio. This makes sense since the Bayesian framework has no linear assumption and, in principle, is applicable to nonlinear systems.   
  }

\vspace{-0.3cm}
\begin{table}[htbp]
  \centering 
  \caption{Validation of the Proposed Method under Different R/X Ratios}
  \label{RXratio}
  \begin{tabularx}{\linewidth}{*{13}{p{.10\linewidth}}}
  \hline 
    \multicolumn{7}{c}{ } \\
    \vspace{-0.2cm}
{Samples}   &  {Iterations}  &   {Meter\newline Std. Dev.}  & Forecast\newline Std. Dev. &  Meter\newline Ratio  & { R/X \newline Ratio} & $\rho_{2{\text{nd}}}$ \newline $[\%]$  
\\ \hline 
  $1,000$  &  $6$ & $1\%$ &  $5\%$  &  $30\%$   &  $\times 1.0$  &  $95.2$ \\
  $1,000$  &  $6$ & $1\%$ &  $5\%$  &  $30\%$   &  $\times 1.5$  &  $93.9$ \\
 $1,000$  &  $6$ & $1\%$ &  $5\%$  &  $30\%$   &  $\times 2.0$ &  $91.7$ \\
\bottomrule
  \end{tabularx}

\end{table}

\subsection{{Further Discussions}}

\subsubsection{{Discussions on Parameter Tuning}} 
{In general, parameter tuning is almost an inevitable task for the statistical-inference-based algorithm. The same story applies to our proposed AIS algorithm as well. In our method, the tunable parameters mainly include: (i) the upper and lower bounds for $p_\text{bino}$, (ii) the iteration number, $j_{\max}$, and (iii) the sample size for each iteration. In this paper, we have conducted extensive case studies that reach the following conclusions for the tuning process of each parameter. As shown in Table~\ref{bounds}, the upper bounds cannot be set to a number very close to $1$ to ensure the algorithm's capability to better search for the global optimal in each iterations. Also, as shown in Table~\ref{ieee1282}, we only need a small number of iterations (e.g., $2$) for the IEEE 123-bus system and $6$ for the 1282-bus system, and a reasonable sample size, e.g. $1,000$, to attain a good estimation accuracy while enabling fast computation for online applications.}

\subsubsection{{Statistical Inference versus Optimization}}
{
Let us now compare the statistical-inference-based algorithm (see, e.g., our work as well as \cite{zhao2019learning} and \cite{singh2010recursive}) to the optimization-based method (see, e.g., \cite{gandluru2019joint} and \cite{cavraro2019real}) in the topology estimation problem. In general, the optimization-based method can directly formulate the topology estimation problem into a mixed-integer program that can be efficiently solved through some packages or commercial software. In general, it demonstrates a good estimation accuracy and a higher computing efficiency than a statistical-inference-based algorithm that  relies on the sampling procedure. However, the statistical-inference-based algorithm also has its own benefits. Unlike the optimization method that only provides a detailed value for the estimation result, the statistical-inference-based algorithm  also provides a confidence interval of the solutions. To illustrate, some of the switch statuses are incorrectly estimated in our case; if the values of the estimated posterior $\hat{\bm{B}}$ are close to $0.5$, we will place less confidence on the estimation results, which are useful information in practice.}

\subsubsection{{Discussions on Observability}}
{Here, we would like to emphasize that although we randomly select the locations of the meters without using an advanced meter placement strategy, our algorithm already demonstrates quite good estimation accuracy as shown in Group 3 in Table~\ref{ieee1282}. We also believe that if proper sensor placement strategies (e.g., in \cite{cavraro2019real,sevlian2017outage,baldwin1993power,gou2008generalized, xygkis2016fisher,liu2014optimal,liu2012trade}) are adopted, the performance of the proposed method still has the potential to be further improved. Also, since the meter placement strategy is not the focus of this paper, we will not initiate further discussion on it.} 

{Also, it is worth pointing out that although we simplify the observability problem to only compare the meter-ratio index of the end-users as shown in Table~\ref{ieee1282}, the actual problem  is much more complicated for the following reasons:
}
\begin{itemize}
\item {First, the observability analysis is \emph{problem-dependent}. It is related to not just intrinsic properties of different test systems (e.g., the size, structure, etc.), but the aforementioned meter locations. Further, different events or outages can give rise to different system topologies that also have impacts on the system topology.  }
\item {Second, while we do not consider the high-renewable-penetrated distribution system in our work, its observability analysis may become even more challenging if the uncertainties brought by the stochastic nature of renewables are considered. Indeed, in our recent research on the observability analysis for a stochastic system, we realize that the traditional deterministic-technique-based observability analysis tool has some limitations in quantifying the observability of a stochastic power system, which exhibits  more complicated phenomena, e.g., \emph{puny} and \emph{brawny} observability phenomena addressed in our recent work \cite{zheng2021observability,zheng2021derivative}. This is also addressed in \cite{augusto2016probabilistic}.
}
\end{itemize}
{Therefore, the observability analysis in distribution system topology estimation is, indeed, a complicated problem that deserves more careful consideration. 
}


\subsubsection{{Discussions on Outlier Issues}} {In practice, there exist three types of outliers (i.e., the observation, innovation, and structure outliers) \cite{gandhi2009robust} that can bias the estimator. Thus, it comes as no surprise that several robust techniques have been proposed. Examples include the $\ell_{1}$-norm estimator \cite{korkali2013robust}, the Huber estimator \cite{mili1999robust}, or more advanced projection-statistic-based generalized
maximum-likelihood-type estimator (known as  the GM estimator) that can better handle the leverage points \cite{gandhi2009robust}.
Moreover, it is important to point out that although we do not address the data asynchronism issue in this paper, the asynchronous data do pollute the measurement quality in practice 
\cite{cavraro2019dynamic}, which can, in turn, bias the estimator. Our proposed Bayesian method is not robust to the aforementioned outliers or bad data as its influence function has not yet been designed to be bounded. Thus, robustifying our proposed Bayesian framework would be a worthwhile future effort.}

\vspace{-0.2cm}
\section{Conclusions and Future Work}
In this paper, we propose an adaptive-importance-sampling-enhanced Bayesian framework to conduct the topology, outage, and state joint estimation with limited measurement devices. Under the validity of the assumptions underlying the proposed Bayesian framework, the bias in the state estimation caused by the pseudomeasurement is canceled in the outage section without the usage of the ping measurement.
By various cases studies conducted in a MATLAB\textsuperscript{\textregistered}-OpenDSS co-simulation environment, the excellent performances of the proposed method are demonstrated in two unbalanced distribution systems.

{As we discussed earlier, the topology estimation problem in practice might be more complicated than the one stated in this paper. To further improve the proposed method's applicability to practical problems, as part of our future work, we will explore the following aspects:}

\begin{itemize}
\item {In practice, the measurements may be corrupted by outliers that can bias the estimators while our current Bayesian estimator has not yet been robustified. Thus, the robustification will be addressed in a future work.}
\item {The loads in practical distribution system can demonstrate non-Gaussian and discrete behaviors that deserve further exploration in the topology estimation problem.}
\item {Observability of a distribution system is a bottleneck for most of the estimation techniques and this is especially true if the penetration of the renewables (e.g., wind and solar) is high since it can greatly increase the uncertainties in the distribution system, which will inevitably affect the accuracy of the estimator. Therefore, we will further develop a strategy to improve the performance of the estimator under high penetration of renewable units.}
\end{itemize}

\vspace{-0.2cm}
\section*{Acknowledgment}

{We would like to appreciate the valuable suggestions and the industrial experiences provided by our colleagues from Dominion Energy and Dr. Hao Huang. 
Furthermore, careful reading and helpful suggestions of the editor and four anonymous reviewers markedly improved the manuscript.}


 \vspace{-0.3cm}
 \newcommand{\BIBdecl}{\setlength{\itemsep}{0.01 em}}
 \bibliographystyle{IEEEtran}
\bibliography{IEEEabrv,References.bib}

\vspace{-1.2cm}
\begin{IEEEbiography}[{\includegraphics[width=0.9in,height=2in,clip,keepaspectratio]{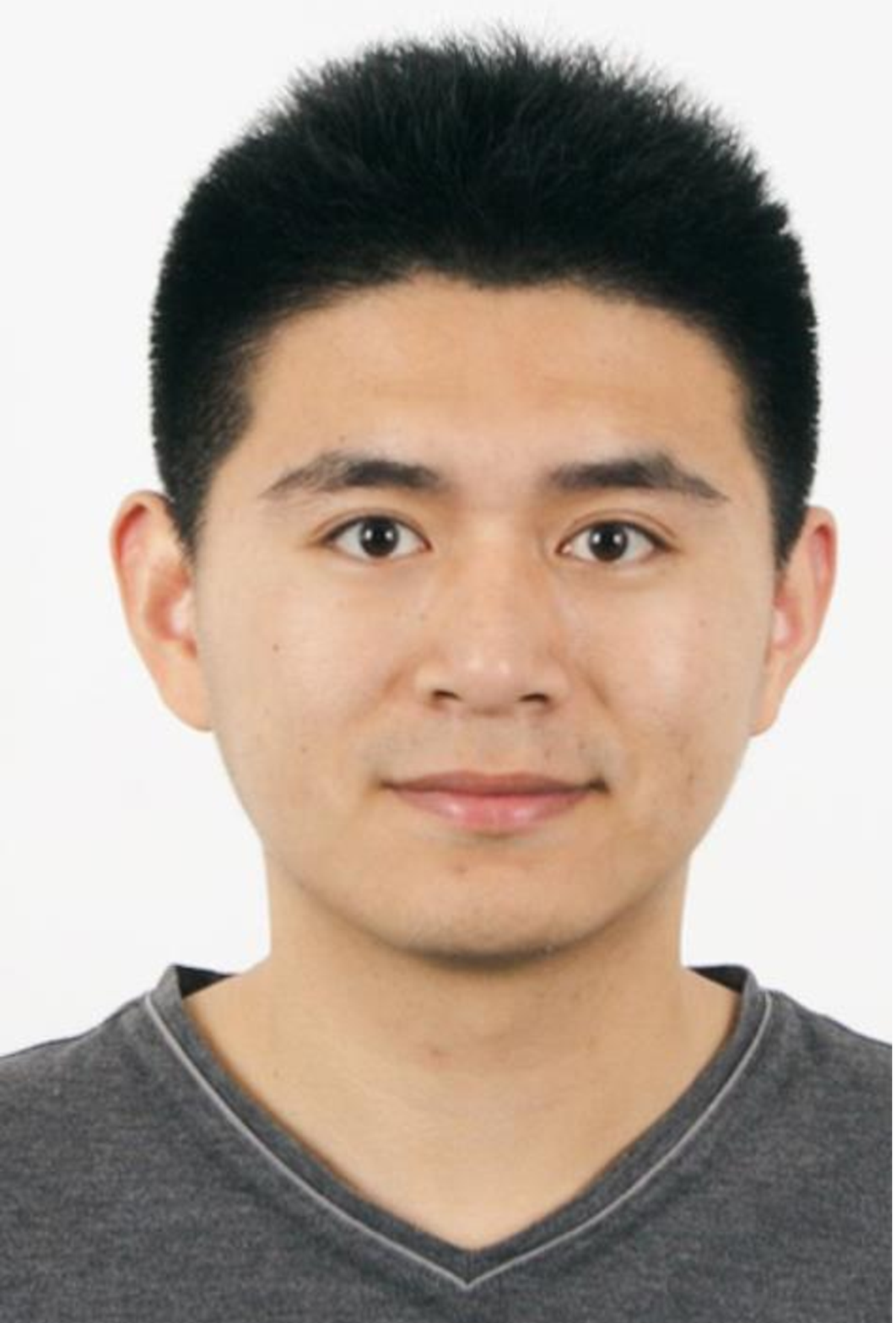}}]{Yijun Xu}(SM'21) received the Ph.D. degree from Bradley Department of Electrical and Computer Engineering at Virginia Tech, Falls Church, VA, on December, 2018. He is currently a research assistant professor at Virginia Tech-Northern Virginia Center, Falls Church, VA. He was a
a postdoc associate at same institute during 2019 to 2020.  
He did the computation internship at Lawrence Livermore  National Laboratory, Livermore, CA, and power engineer internship at ETAP -- Operation Technology, Inc., Irvine, California, in 2018 and 2015, respectively. 

His research interests include power system uncertainty quantification, uncertainty inversion, and decision-making under uncertainty. 
Dr. Xu is currently serving as an Associate Editor of the \textsc{IET Generation, Transmission \& distribution} and an Associate Editor of the \textsc{IET Renewable Power Generation}. He is the co-chair of the IEEE Task Force on Power System Uncertainty Quantification and Uncertainty-Aware Decision-Making.
\end{IEEEbiography}

\begin{IEEEbiography}[{\includegraphics[width=1in,height=3in,clip,keepaspectratio]{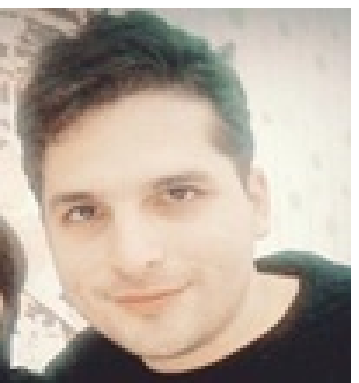}}]{Jaber Valinejad} (M'19) is currently pursuing his Ph.D. degree at the Bradley Department of Electrical and Computer Engineering, Virginia Tech, Greater Washington, D.C., USA. He is also pursuing an MSc degree at the Department of Computer Science at the same school. He is with an NSF-sponsored interdisciplinary disaster resilience Program. His current research interests include power systems, resilience and community resilience, cyber-physical–social systems and social computing, artificial intelligence, and learning. 
\end{IEEEbiography}

\vspace{-0.5cm}
\begin{IEEEbiography}[{\includegraphics[width=1in,height=3in,clip,keepaspectratio]{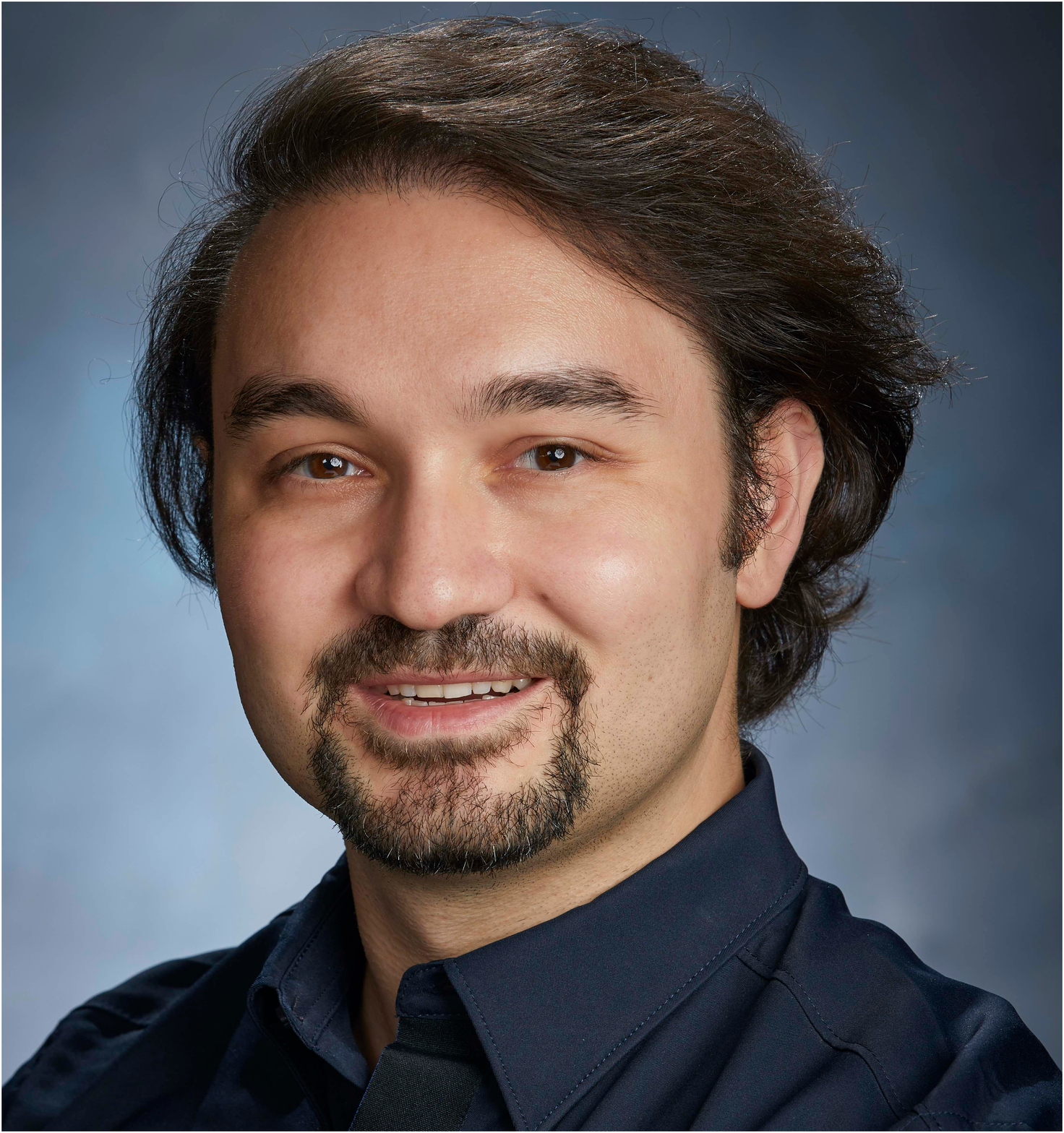}}]{Mert Korkali} (SM'18) received the Ph.D. degree in electrical engineering from Northeastern University, Boston, MA, in 2013. He is currently a Research Staff Member at Lawrence Livermore National Laboratory, Livermore, CA. From 2013 to 2014, he was a Postdoctoral Research Associate at the University of Vermont, Burlington, VT. 

His current research interests lie at the broad interface of robust state estimation and fault location in power systems, extreme event modeling, cascading failures, uncertainty quantification, and probabilistic grid planning. He is the Co-chair of the IEEE Task Force on Standard Test Cases for Power System State Estimation and the Secretary of the IEEE Task Force on Power System Uncertainty Quantification and Uncertainty-Aware Decision-Making. Dr. Korkali is currently serving as an Editor of the \textsc{IEEE Open Access Journal of Power and Energy} and of the \textsc{IEEE Power Engineering Letters}, and an Associate Editor of \emph{Journal of Modern Power Systems and Clean Energy}.
\end{IEEEbiography}

\vspace{-0.3cm}
\begin{IEEEbiography}[{\includegraphics[width=1.0in,height=1.2in,clip,keepaspectratio]{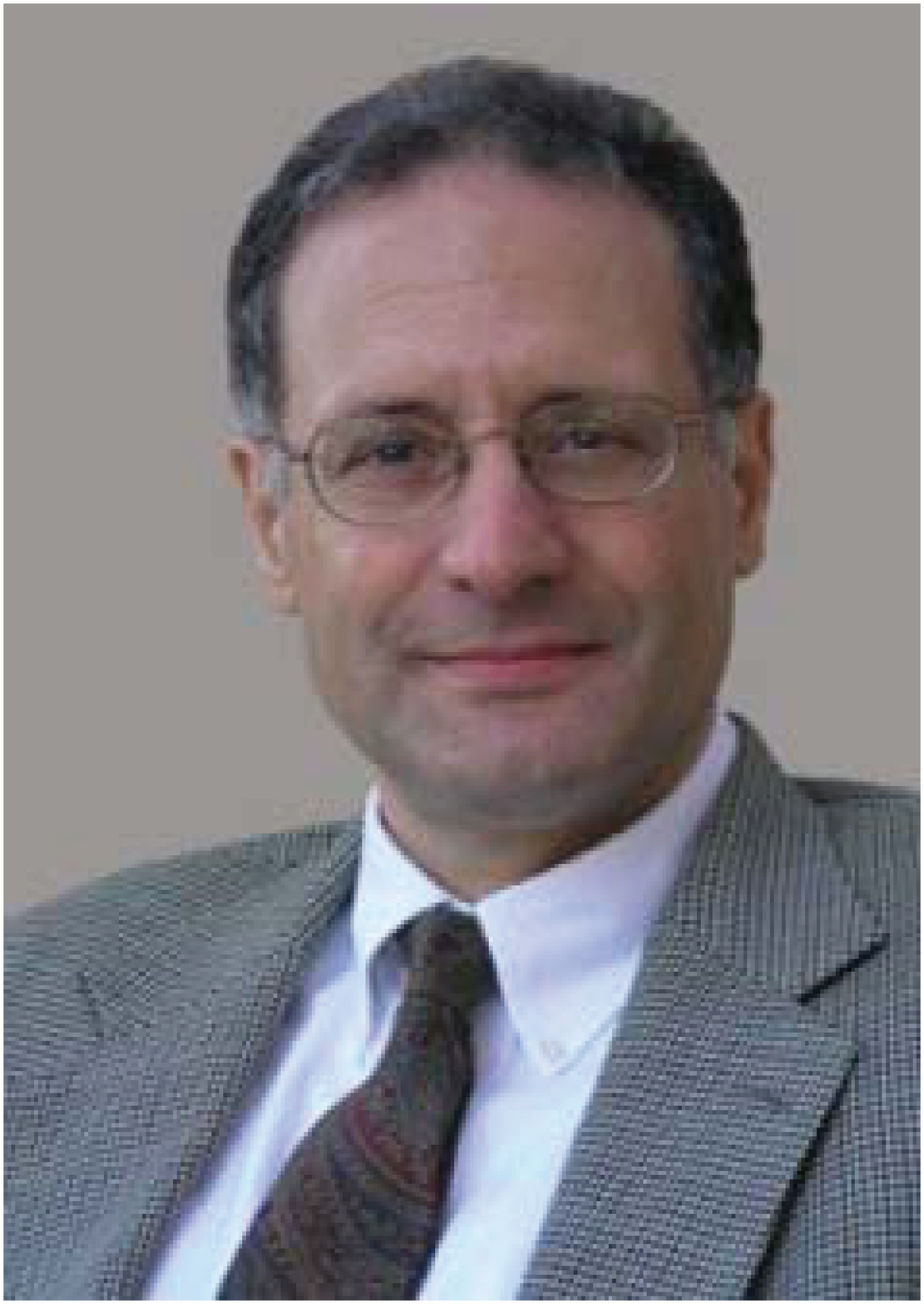}}]{Lamine Mili}(LF'17) received the Ph.D. degree from the University of Li{\`e}ge, Belgium, in 1987. He is a Professor of Electrical and Computer Engineering, Virginia Tech, Blacksburg. He has five years of industrial experience with the Tunisian electric utility, STEG. At STEG, he worked in the planning department from 1976 to 1979 and then at the Test and Meter Laboratory from 1979 till 1981. He was a Visiting Professor with the Swiss Federal Institute of Technology in Lausanne, the Grenoble Institute of Technology, the {\'E}cole Sup{\'e}rieure D'{\'e}lectricit{\'e} in France and the {\'E}cole Polytechnique de Tunisie in Tunisia, and did consulting work for the French Power Transmission company, RTE.

His research has focused on power system planning for enhanced resiliency and sustainability, risk management of complex systems to catastrophic failures, robust estimation and control, nonlinear dynamics, and bifurcation theory. He is the co-founder and co-editor of the \emph{International Journal of Critical Infrastructure}. He is the chairman of the IEEE Working Group on State Estimation Algorithms and the chair of the IEEE Task Force on Power System Uncertainty Quantification and Uncertainty-Aware Decision-Making. He is a recipient of several awards including the US National Science Foundation (NSF) Research Initiation Award and the NSF Young Investigation Award.
\end{IEEEbiography}

\vspace{-0.5cm}
\begin{IEEEbiography}[{\includegraphics[width=1in,height=3in,clip,keepaspectratio]{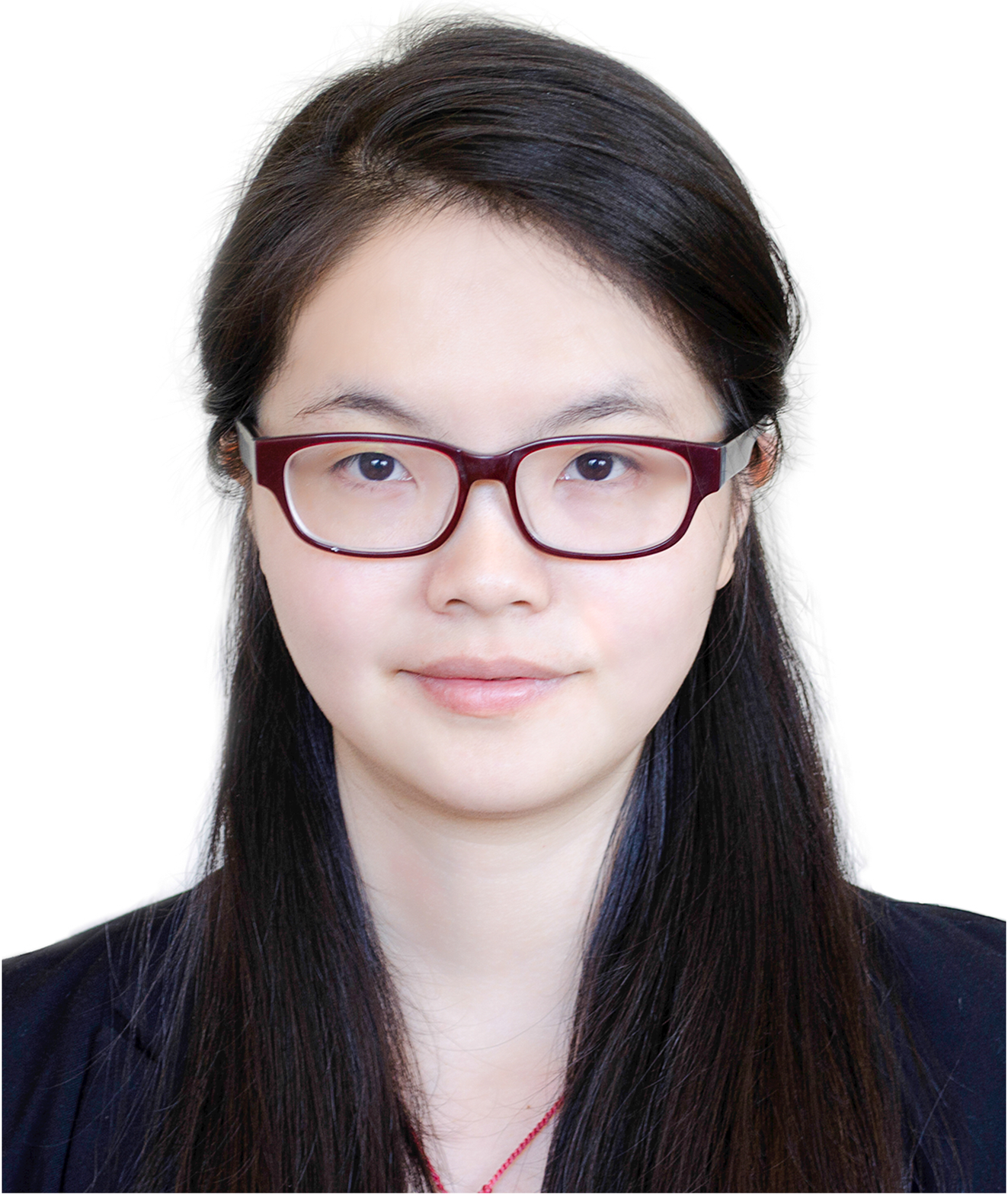}}]{Yajun Wang} (SM'21) received the B.Sc. and M.Sc. degrees from the School of Electrical Engineering, Wuhan University, Wuhan, China, in 2012 and 2014, respectively. She received her Ph.D. degree in Electrical Engineering from the University of Tennessee, Knoxville, TN, in 2019. She is currently working as a Senior Power System Engineer and distribution storage Technical Lead with Dominion Energy Virginia, Richmond, VA. Her research interests are energy storage system, renewable energy integration, vehicle-to-grid, big data analytics in power systems, stability and control, and system restoration.
\end{IEEEbiography}

\vspace{-0.3cm}
\begin{IEEEbiography}[{\includegraphics[width=1.0in,height=1.4in,clip,keepaspectratio]{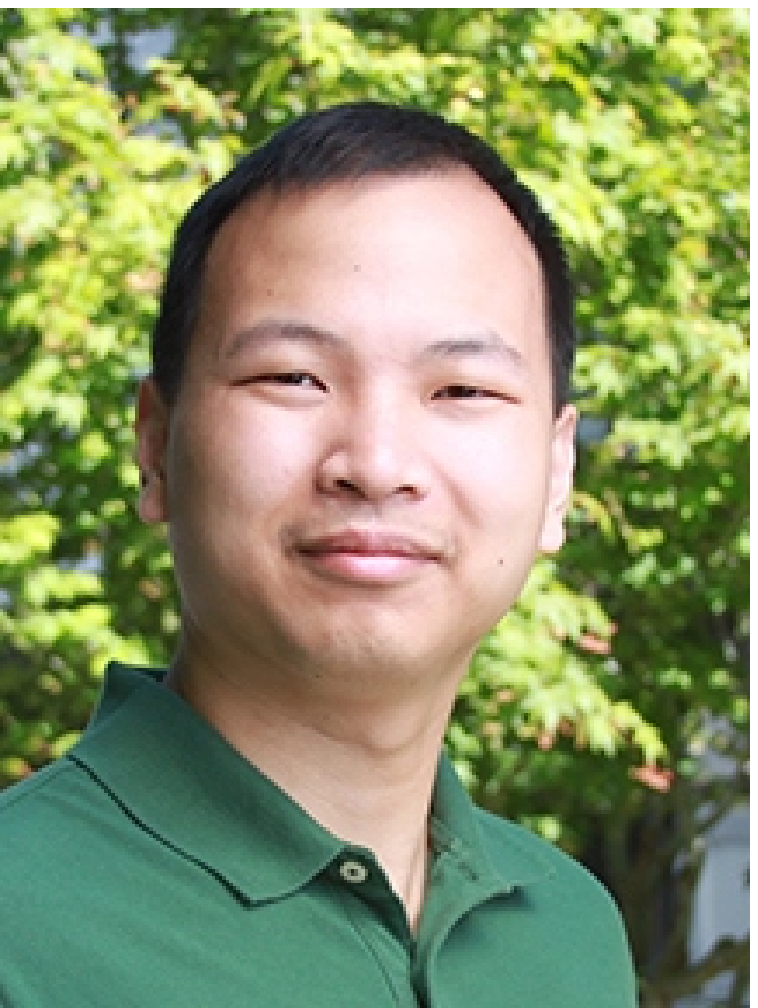}}]{Xiao Chen} received his Ph.D. degree in Applied Mathematics from Florida State University, Tallahassee, FL in 2011. He is a Computational Scientist and a Project Leader in the Center for Applied Scientific Computing at Lawrence Livermore National Laboratory, Livermore, CA. He works primarily on the development and application of advanced computational and statistical methods and techniques to power engineering, reservoir simulation, subsurface engineering, and seismic inversion. His research interests include uncertainty quantification, data assimilation, and machine learning. 
\end{IEEEbiography}

\begin{IEEEbiography}[{\includegraphics[width=1in,height=1.25in,clip,keepaspectratio]{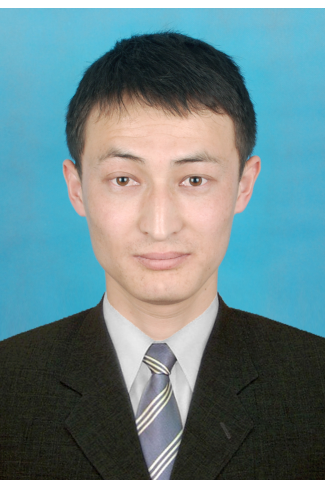}}]
{Zongsheng Zheng} (M'20) received the Ph.D. degree in electrical engineering from Southwest Jiaotong University, Chengdu, China, in 2020. During 2018-2019, he was a Visiting Scholar at the Bradley Department of Electrical and Computer Engineering at Virginia Tech-Northern Virginia Center, Falls Church, VA, USA. He is currently an Research Associate Professor at the College of Electrical Engineering, Sichuan University. His research interests include uncertainty quantification, parameter and state estimation.
\end{IEEEbiography}

\end{document}